\documentclass[preprint]{aastex}

\begin{document}

\title{The Tolman Surface Brightness Test for the Reality of the Expansion.
 IV. A Measurement of the Tolman Signal and the Luminosity Evolution of
 Early-Type Galaxies}

\author{Lori M. Lubin\altaffilmark{1,2}} \affil{Department of Astronomy,
California Institute of Technology,\\ Mailstop 105-24, Pasadena,
California 91125}

\author{Allan Sandage} 
\affil{Observatories of the Carnegie Institution of Washington,\\ 813
Santa Barbara Street, Pasadena, California 91101}

\altaffiltext{1}{Hubble Fellow} 
\altaffiltext{2}{Current address : Department of Physics and
Astronomy, Johns Hopkins University, Baltimore, MD 21218}

\begin{abstract}

We review a sample of the early literature in which the reality of the
expansion is discussed. Hubble's reticence, even as late as 1953, to
accept the expansion as real is explained as due to his use of
equations for distances and absolute magnitudes of redshifted galaxies
that do not conform to the modern Mattig equations of the standard
model. The Tolman surface brightness test, once the only known test
for the reality of the expansion, is contrasted with three other
modern tests.  These are (1) the time dilation in Type Ia supernovae
light curves, (2) the temperature of the relic radiation as a function
of redshift, and (3) the surface brightness normalization of the
Planckian shape of the relic radiation.

We search for the Tolman surface brightness depression with redshift
using the {\it Hubble Space Telescope} (HST) data from Paper III for
34 early-type galaxies from the three clusters Cl 1324+3011 ($z =
0.76$), Cl 1604+4304 ($z = 0.90$), and Cl 1604+4321 ($z =
0.92$). Depressions of the surface brightness relative to the
zero-redshift fiducial lines in the mean surface brightness, log
linear radius diagrams of Paper I are found for all three
clusters. Expressed as the exponent, $n$, in 2.5 log $(1 + z)^n$ mag,
the value of $n$ averaged over Petrosian radii of $\eta = 1.7$ and
$\eta = 2.0$ for all three clusters is $n = 2.59 \pm 0.17$ in the $R$
band and $3.37 \pm 0.13$ in the $I$ band for a $q_o = {1\over{2}}$
model.  The sensitivity of the result to the assumed value of $q_o$ is
shown to be less than 23\% between $q_o = 0$ and $+1$.  The conclusion
is that the exponent on $(1 + z)$ varies from 2.28 to $2.81~(\pm
0.17)$ in the $R$ band and 3.06 to $3.55~(\pm 0.13)$ in the $I$ band,
depending on the value of $q_o$. For a true Tolman signal with $n =
4$, the luminosity evolution in the look-back time, expressed as the
exponent in 2.5 log $(1 + z)^{4-n}$ mag, must then be between 1.72 to
$1.19~(\pm 0.17)$ in the $R$ band and 0.94 to $0.45~(\pm 0.13)$ in the
$I$ band. We show that this is precisely the range expected from the
evolutionary models of Bruzual \& Charlot (1993, ApJ, 405, 538) and
other measurements of the luminosity evolution of early-type
galaxies. We conclude that the Tolman surface brightness test is
consistent with the reality of the expansion to within the combined
errors of the observed $\langle SB \rangle$ depression and the
theoretical correction for luminosity evolution.

We have also used the high-redshift HST data to test the ``tired
light'' speculation for a non-expansion model for the redshift. The
HST data rule out the ``tired light'' model at a significance level of
better than 10 $\sigma$.

\end{abstract}

\keywords{galaxies: clusters: general -- cosmology: observations}

\section{Introduction}

\subsection{Early Commentaries on the Reality of the Expansion}

With the announcement by Hubble (1929a) of a correlation between his
estimates of distances to nearby galaxies and their redshifts,
observational cosmology came of age, beginning its Long Journey into
Night. The redshifts used by Hubble had been measured by Slipher,
published by Eddington (1923), and added to by Humason (1929) in his
crucial extension to higher values before Hubble's announcement.

However, the announcement of an expanding universe was such an
extraordinary claim that proof of its reality by some independent
means seemed essential, even though an expanding universe had been
predicted by Friedmann (1922) [cf.\ Tropp, Frenkel \& Chernin 1993] as
one of the solutions of fundamental Einstein equation of general
relativity.  Expanding solutions were also detailed later by Lemaitre
(1927, 1931) and Robertson (1928), each of which, acknowledging
Friedmann, made advances beyond Friedmann by their adumbrations
concerning the available observations of galaxies and their relative
distances.

It is not clear whether Hubble knew of the Friedmann (1922) prediction
or of the theoretical, cum observational, papers of Lemaitre and
Robertson in 1927 and 1928, although Robertson told one of us (AS)
that he had discussed with Hubble the existence of an expanding
solution to the Einstein equations before 1929 (e.g.\ Sandage 1995,
footnote 16 to Chapter 5). None of these three principal theoreticians
are mentioned in Hubble's 1929 announcement.

Nevertheless, the decade of the 1920's was not entirely free of
observational attempts to test a related cosmological prediction. A
year after the publication of Einstein's field equations, de Sitter
(1917a,b) had discovered one curious solution to them. Although his
solution was that of a static metric (there are only three such
solutions; Tolman 1929), it remarkably exhibited a redshift. The
metric coefficient of the four-space time coordinate was a function of
distance from the observer. Clocks that are further from the
observer's origin in the three-space manifold would appear to tick
more slowly than clocks at that origin. This effect would give an
apparent redshift that would vary with distance. The formal feature of
a redshift in the de Sitter metric was called ``the de Sitter
effect.''  A curious additional feature was that any test particle put
into the de Sitter manifold would exhibit a radial motion (Eddington
1923, \S 70; de Sitter 1933; Tolman 1934, \S 144--145)\footnote{An
accessible derivation of the de Sitter static metric and its
properties is given by Tolman (1934, \S 136 and \S 142--145). A more
recent summary of the de Sitter metric with its redshift properties is
given elsewhere (Sandage 1975, \S 2.2).} even though the spatial
metric was static.

The predicted existence of the de Sitter redshift effect with its
static metric became well known in the decade of the 1920's. Many
attempts were made to find the effect using astronomical data
(distances and velocities) for objects such as stars and globular
clusters, thought then to comprise the wider universe before Hubble
(1925, 1926, 1929b) proved the existence of external galaxies. Among
the most accessible papers concerning the search for the de Sitter
effect are those by Silberstein (1924), Stromberg (1925), Wirtz
(1925), Lundmark (1925), and undoubtedly many others.

With regard to the de Sitter effect, it is a continuing curiosity as
to what Hubble meant in the final sentence of his announcement in 1929
where he wrote:

\begin{quote}
The outstanding feature is the possibility that the velocity-distance
relation may represent the de Sitter effect.... In this connection it
may be emphasized that the linear relation found in the present
discussion is a first approximation {\it representing a restricted
range in distance} (emphasis added).
\end{quote}

In other words, a larger range in distance may not show a linear
relation, he surmised, where he apparently knew that the first order
de Sitter effect is {\it quadratic} in distance rather than linear
(e.g.\ Sandage 1995, eq.\ 5.10). Cautiously, Hubble left open the
possibility that the data which he had discussed might define a
redshift-distance relation that would actually vary as the square of
the distance, which, in first approximation, would appear to be linear
for short distances near the origin (i.e. the first term of a Taylor
series). Hubble also undoubtedly knew of the earlier searches for the
de Sitter effect by Lundmark (1925) and Silberstein (1924) where they
attempted parabolic fits to their adopted data.

However, Hubble \& Humason (1931) soon proved that the
redshift-distance relation was, in fact, linear over the much larger
range of redshifts than was available in 1929. For this and other
reasons, de Sitter (1933) wrote: ``We know now, because of the
observed expansion, that the actual universe must correspond to one of
the non-static models.... The static models are, so to say, only of
academic interest.''
 
\subsection{Later Attempts to Counter the Reality of the Expansion} 
     
Nevertheless, the concept of an expanding universe seemed so bizarre
to many commentators that attempts began already in 1929 to find
alternate ways to produce large redshifts other than from a true
expansion. These attempts continue to this day.

The first alternate suggestion was made by Zwicky (1929) where he
proposed that photons lose energy on their way to us from a distant
source. As a consequence, they would show a redshift in their energy
distribution, both in the continuum radiation and in the Fraunhofer
lines used to measure the redshifts. To first order, the redshift
effect would be linear with distance because the first term in a
Taylor expansion of $1 + z = e^{{HD}\over{c}}$ is linear in $D$. Here,
$H$ is the Hubble constant and $D$ is the distance. With this
suggestion, Zwicky (1929) introduced the notion of ``tired light.''

Even as late as mid-twentieth century, Zwicky (1957) maintained that
the hypothesis was viable. However, neither Zwicky nor any other
subsequent supporter of the proposition (e.g.\ La Violette 1986;
Pecker \& Vigier 1987) gave a convincing physical theory for the
``tiredness.'' As critics still point out, any scattering process with
energy transfer from the photon beam to the scattering medium, as
required for a redshift, must broaden (deflect) the beam. This effect
would cause images of distant galaxies to be fuzzier than their local
counterparts, which they are not.

\subsection{Hubble's Reticence}

Perhaps the most interesting attack on the reality of the expansion
was the reluctance of Hubble himself to believe that the redshifts
represent a true expansion rather than ``caused by an unknown law of
nature.'' Much has been written about Hubble's reluctance, most of
which is wrong. Some commentators even suggest philosophical or
religious reasons related to a presumed abhorrence of a ``creation''
event that is implied in some interpretations of a real Friedmann
expansion.

However, the fact is that Hubble's reasons were those of a
reductionist bench scientist. He relied (mistakenly, it turns out)
solely on the interpretation of his observational data and their
accuracy, coupled with a mistaken theory of how redshifts should vary
with ``distance.''  His equation for ``distance'' has no justification
within the modern Mattig (1958) equations. This story, and why
Hubble's conclusion would not have been reached using current data
analyzed with the modern theoretical equations, is set out in detail
elsewhere (Sandage 1998).

In outline, Hubble's argument was as follows. By the mid 1930's,
Hubble (1934; 1936a,b) had completed his program of galaxy counts, the
goal of which was to measure the curvature of space.  He also
completed the extension of the Hubble diagram of redshift versus
apparent magnitude to the limit of the Mount Wilson 100-inch reflector
(Hubble 1936; Humason 1936).  The data for each of these programs had
to be corrected for the effects of redshifts on the apparent
magnitudes. If the expansion was real, Hubble assumed that the
observed bolometric magnitudes (or equivalently, the observed
magnitudes corrected for the selective part of the $K$ term) must be
made brighter by two factors of 2.5 log $(1 + z)$; however, if the
redshift was due to ``an unknown law of nature'' rather than the
expansion, only one such factor was to be applied to the observed
magnitudes.

Hubble believed that the Hubble diagram (log of the redshift versus
apparent magnitude) must be strictly linear; in addition, the radius
of curvature of space implied by his ``corrected'' magnitudes for his
galaxy counts must not be ``too small.''  Consequently, he became
convinced that only one factor of 2.5 log $(1 + z)$ should be
applied. If so, the expansion would not be real. Hubble (1936b) wrote:

\begin{quote}

If the redshifts are not primarily due to velocity shifts ... [then]
the velocity-distance relation is linear; the distribution of nebulae
is uniform; there is no evidence of expansion, no trace of curvature,
no restriction of the time scale....  The unexpected and truly
remarkable features are introduced by the additional assumption that
redshifts measure recession. The velocity-distance relation deviates
from linearity by the exact amount of the postulated recession. The
distribution departs from uniformity by the exact amount of the
recession.  The departures are compensated by curvature which is the
exact equivalent of the recession. Unless the coincidences are
evidence of an underlying necessary relation between the various
factors, they detract materially from the plausibility of the
interpretation. The small scale of the expanding model, both in space
and time, is a novelty and as such will require rather decisive
evidence for its acceptance.

\end{quote}
   
\noindent That ``rather decisive evidence'' is now available from at
least three modern experiments that are independent of the Tolman
galaxy surface brightness test.

\subsection{Other Recent Proofs that the Expansion is Real}
              
\subsubsection{The Time Dilation Test}

The Tolman (1930) surface brightness $(1 + z)^4$ effect was the only
known test for the reality of the expansion until Wilson (1939)
suggested that the shape of the light curves of Type Ia supernovae
provides a clock. This supposition was based on the uniform shape of
the light curve discovered by Baade (1938) and recalled as history by
Minkowski (1964). Wilson reasoned that such clocks at different
redshifts would measure the special relativity time dilation if the
light curve shapes of SNe Ia at high redshifts could be observed. The
stretching of the light curves, increasing with redshift, has now been
observed. The data give a spectacular confirmation of the time
dilation effect (Goldhaber et al.\ 1997, 2001).

\subsubsection{Temperature of the Relic Blackbody Radiation as a 
Function of Redshift}

Blackbody radiation in an expanding cavity remains Planckian in shape
but with a decreasing temperature that scales as $T(z) = T(0)(1+z)$
with redshift (Tolman 1934, eq. 177.7; Bahcall \& Wolf 1968). The
observations of the Boltzman temperature of interstellar molecules in
the spectra of high-redshift galaxies has now apparently been measured
in a difficult experiment with the Keck 10-m telescopes (Songaila et
al. 1994).\ An important confirmation of these first results was made
in observations by Ge et al.\ (1997) and Srianand, Petitjean \& Ledoux
(2000). Only upper limits had been achieved before (cf.\ Meyer et al.\
1986) which, nevertheless, were highly important in pioneering this
test.

\subsubsection{Measurement of the Chemical Potential the Alpher-Herman 
Relic Black Body Radiation}
     
Although the {\it shape} of an initial blackbody spectrum remains
Planckian in an expanding cavity, the vertical normalization (i.e.
the photon number) remains Planckian only if that normalization is
decreased with redshift by $(1+z)^4$. This fact is derived trivially
from the Planck equation as expressed in terms of energy flux per
wavelength per unit wavelength interval rather than frequency per unit
frequency interval. (Of course, both representations are equivalent by
minding the relation between wavelength interval and frequency
interval, where $d \nu = c {{d\lambda}\over {\lambda^2}}$.)

Hence, because the Planck equation defines a {\it surface brightness},
a test of the Tolman surface brightness effect is equivalent to
measuring the deviation of the photon number per unit surface area in
the sky by comparing the observations with the normalization given by
the Planck equation itself. The deviation of the data from the Planck
equation is called the ``chemical potential.'' Among other things, the
deviation with wavelength could be due to Compton scattering in the
early universe.

No deviation has been found in the observations to within one part in
$10^4$. The perfect Planckian shape of the relic radiation was
measured with COBE to within this limit of $9 \times 10^{-5}$ (Mather
et al. 1990; Fixsen et al.\ 1996). The conclusion to be drawn from
this spectacular result is that this seemingly perfect normalization
of the spectral energy distribution is a definitive proof of the
Tolman surface brightness factor and, therefore, a definitive proof of
the reality of the expansion. We shall adopt this assumption later in
\S 4 where we combine our surface brightness signal with the
theoretical $(1 + z)^4$ Tolman signal and interpret the difference to
be due to luminosity evolution in the high-redshift look-back time.

\subsection{Plan of the Present Paper}

In the first three papers of this series (Sandage \& Lubin 2001,
hereafter Paper I; Lubin \& Sandage 2001a and 2001b, hereafter Papers
II and III), we provide the background and observational data required
to carry out the Tolman test.  Most importantly, we measure the
fiducial (zero-redshift) relations between the mean surface
brightness, absolute magnitude, and linear radius for local early-type
galaxies in Paper I. In Paper III, we present the observational data
on our high-redshift comparison sample of 34 early-type galaxies in
the three clusters used in this program, Cl 1324+3011 at $z = 0.76$,
Cl 1604+4304 at $z = 0.90$, and Cl 1604+4321 at $z = 0.92$ (Oke,
Postman \& Lubin 1998; Postman et al.\ 1998, 2001; Lubin et al.\ 1998,
2001).  To compare accurately the local and high-redshift data, the
parameters of each galaxy are measured at discrete values of the
Petrosian (1976) $\eta$ metric radius, which is defined as the
difference in magnitude between the mean surface brightness averaged
over the area interior to a particular radius and the surface
brightness at that radius (see \S2 of Paper I).

In the present paper, we use the results presented in Papers I--III to
complete the Tolman test. In \S 2 we review the Mattig (1958)
cosmological equations with which to calculate the absolute magnitudes
and linear radii of galaxies from their apparent magnitudes and
angular radii. Explicit equations for the special case of $q_o =
{1\over{2}}$ are given. We use these equations to obtain the total
magnitude $M$, linear radius R, and mean surface brightness $\langle
SB \rangle$ as functions of five Petrosian $\eta$ radii of 1.0, 1.3,
1.5, 1.7, and 2.0 mag for the 34 high-redshift early-type galaxies.
Tables 2--4 list these values for $q_o = {1\over{2}}$ and $H_o = 50$
km s$^{-1}$ Mpc$^{-1}$.  These values are derived from the
observational data listed in Paper III.  The Tolman test is made in \S
3 based on the data in \S 2 and the comparison with the surface
brightness data at zero redshift from Paper I.

In \S 4 we set out the theoretical evolutionary corrections, first,
using the simplest model of passive evolution by main sequence
burn-down in the HR diagram as a function of time and, second, using
the sophisticated star-formation models of Bruzual \& Charlot (1993),
following earlier papers by Guiderdoni \& Rocca-Volmerange (1987,
1988) and Rocca-Volmerange \& Guiderdoni (1988).

In \S 5 we show the sensitivity of the observed Tolman signal to the
assumed value of $q_o = {1\over{2}}$ used in the data of Tables 2--4.
Proof that the ``tired light'' assumption does not fit the data by a
large factor is given in \S 6. In \S 7, we discuss the systematic
uncertainties in the Tolman test made here and describe plans to
strengthen the present test.

\section{Calculation of Absolute Magnitudes and Linear Radii for the 
High-Redshift Galaxies}

Comparison of the high-redshift data in Paper III with the data for
local galaxies in Paper I requires knowledge of absolute magnitudes
(corrected for $K$ term) and linear radii.

We have two options to calculate these values. (1) If we were to
assume that the standard model that leads to the Mattig (1958, 1959)
equations (which assume that the expansion is real) did not exist, the
natural assumption is that the distance, $D_o$, at the time that light
is received is related to redshift by $D_o = {{cz}\over{H_o}}$.  This
was Hubble's assumption throughout his work, including the last
summary paper in his Darwin lecture (Hubble 1953). (2) Alternately, we
can adopt the details of the standard model by choosing a value for
the deceleration parameter, $q_o$ (Hoyle \& Sandage 1956). We then use
the Mattig (1958) equations to calculate the distance modulii, $m -
M$, and the linear radii, R, from the observed angular
radii.\footnote{To avoid confusion with our use of R in this paper to
mean the linear radius of a galaxy rather than its use as the
time-dependent scale factor in the metric, we have used $D$ here as
the distance parameter by replacing $Rr$ by $D$, where, in the
standard notation, $R$ is the scale factor that varies with time and
$r$ is the dimensionless co-moving radial coordinate in the
metric. Hence, in our notation here, $D_o$ is the distance at the time
light is received from a galaxy at redshift $z$. This is given by
Mattig (1958; see also Sandage 1988, eq.\ 30) as
\begin{equation}
D_o = {c\over{H_o q_o^2 (1+z)}} [z q_o + (q_o - 1) \{-1 +
\sqrt{2 q_o z + 1}\}]
\end{equation}} 

Because these equations already contain the Tolman $(1 + z)^4$ factor,
the test for the Tolman effect becomes one of consistency between the
surface brightness observations and the predictions of the standard
model. The argument of why this does not lead to a hermenuetical
circularity was given in \S 5 of Paper I.  We adopt option (2) in
calculating the distance moduli and the linear radii of the program
clusters. We treat only the $q_o = {1\over{2}}$ case of the standard
model in this section. The cases for $q_o = 0$ and $+1$ are treated in
\S 5 using the correction recipes in Table 8.

\subsection{The Equations for $q_o = {1\over{2}}$}
      
For the $q_o = {1\over{2}}$ case with $H_o = 50$ km s$^{-1}$
Mpc$^{-1}$, the Mattig equations reduce to

\begin{equation}
m - M = 5~{\rm log}~[2(1+z - \sqrt{1+z})] + 43.89,
\end{equation}

\noindent and 

\begin{equation}
D_1 = {{2c}\over{H_o(1+z)}}\left [1 - {1\over{\sqrt{1+z}}}\right],
\end{equation}

\noindent where $D_1$ is the distance (in parsecs) when light left the
galaxy at the observed redshift $z$. The linear radius R, in parsecs,
of a galaxy with angular radius $\theta$, in radians, is

\begin{equation}
        {\rm R(pc)} = {{2c~(\sqrt{1+z}-1)}\over{H_o(1+z)^{3\over{2}}}}~\theta
        ({\rm radians}),
\end{equation}

\noindent using the distance $D_1$ from equation (3).

Changing the angular radius in radians to arcseconds, equation (4) can
be conveniently written as

\begin{equation}
{\rm log~R(pc)} = {\rm log}~\theta'' + {\rm log}~A,
\end{equation}

\noindent where 

\begin{equation}
A = 5.818 \times 10^4 \left [{\sqrt{1+z} - 1}\over{(1+z)^{3\over{2}}}
\right ],
\end{equation}

\noindent for $q_o = {1\over{2}}$ and $H_o = 50$. 

Equations 2--4 are derived in Sandage (1961a,b; 1988; 1995), where the
standard model is summarized from the point of view of practical
cosmology.

Table 1 shows the $m - M$ and $A$ values for the three program
clusters for the $q_o = {1\over{2}}$ case using equations (2), (5),
and (6).  The $K$ terms are those derived in Paper III (see Table 4).

\subsection{The $M$, $R$, and $\langle SB \rangle$ Data for the Three 
Program Clusters for the $q_o = {1\over{2}}$ Case}

The results for absolute magnitude $M$, linear radius $R$, and mean
surface brightness $\langle SB \rangle$ are listed in Tables 2--4
where the $\langle SB \rangle$ values are those in Paper III,
corrected for the effects of redshift by the $K$ terms from Table 1.
The format of Tables 2--4 is the same as for Tables 5--7 in Paper
III. Data for the five $\eta$ values of 1.0, 1.3, 1.5, 1.7, and 2.0
are listed. For each $\eta$, the $K$-corrected absolute magnitude is
calculated using the $m - M$ value from Table 1 and the apparent
magnitudes from Paper III. The $K$-corrected mean surface brightnesses
from Paper III are given next. The logs of the linear radii (in
parsecs) are obtained by adding the log of the angular radii from
Tables 5--7 of Paper III to the $A$ values from Table 1, calculated
using equations (5) and (6).

\section{The Tolman Test Using the HST Data for Early-Type Galaxies from
the Three Program Clusters for $q_o = {1\over{2}}$}

The data that are necessary to make the Tolman test are given in
Tables 2--4. The search for a Tolman signal, as modified by luminosity
evolution, is made by comparing these high-redshift data with the
calibrations at zero redshift in Paper I. The comparisons for the $R$
photometric band of cluster Cl 1604+4321 ($z = 0.9243$) can be made
directly because the Postman \& Lauer (1995) calibrations in Paper I
are already in the standard Cape/Cousins $R$ photometric band.

Comparisons in the $I$ band for clusters Cl 1324+3011 ($z = 0.7565$)
and Cl 1604+4304 ($z = 0.8967$) are made by converting the
calibrations in Paper I to the $I$ photometric band by applying the
adopted color of $\langle R-I\rangle = 0.62$. This mean ($R-I$) color
for early-type galaxies at zero redshift was derived by (1) using
$\langle B-R \rangle = 1.52$ from the photometry of Postman \& Lauer
(1995), as corrected for $K$ term and Galactic absorption in Paper I,
and (2) using the two-color $(B-R)$, $(R-I)$ diagram from the data of
Poulain \& Nieto (1994) to give $\langle R-I \rangle = 0.62$ at
$\langle B-R \rangle = 1.52$. A color of $\langle R-I \rangle = 0.62$
corresponds to a mean stellar spectral type of about K5 (see Figure 10
of Sandage 1997).
          
\subsection{The Tolman Signal in the log R, $\langle SB \rangle$ 
Correlation Diagram}

We approach the problem first by considering the data in the parameter
space which contains the smallest evolutionary signal and, therefore,
is expected to have the strongest Tolman signal (if it exists). Beyond
a doubt, luminosity evolution occurs in a stellar population with no
new star formation after the first initial star burst. The main
sequence termination point in the Hertzsprung--Russell (HR) diagram
burns down to fainter luminosities as the population ages, making the
total luminosity of the aggregate fainter with time. Calculation of
the burn-down rate, coupled with the luminosity function that
determines relative numbers of stars that partake in the evolution,
gives a first estimate of the change of total luminosity with
time. This is how we make a first estimate of the effect in \S 4.

Luminosity evolution affects both the observed surface brightness and
the absolute magnitude. However, we expect that it does not affect the
radius as a function of time. The consequence is that the correlation
of surface brightness versus absolute magnitude has a double dose of
evolution because both coordinates are affected, whereas the
correlation of surface brightness and linear radius contains only a
single dose, on the assumption that the radius does not change.  On
this basis, the most powerful of the three correlation diagrams from
Paper I is the $\langle SB \rangle$, log R diagram. In this section we
examine this correlation, searching for a Tolman signal in each of the
three high-redshift clusters.

\subsubsection{The data in $R$ for Cl 1604+4321 ($z = 0.9243$)}

Figure 1 shows the $\langle SB \rangle$, log $R$ correlation diagram
for Cl 1604+4321 in the $R$ photometric band as corrected for $K$
dimming using Table 1. Large dots are those galaxies in Table 2 that
have directly measured redshifts. Small dots are assumed cluster
members on the basis of their photometric redshifts (for details, see
Paper III). 

The lines in each panel are the correlations for zero redshift from
Tables 2 and 3 of Paper I based on the photometry by Postman \& Lauer
(1995) and extended to radii smaller than log $R = 4.3$ using the data
of Sandage \& Perelmuter (1991). Note that Table 3 in Paper I lists
the magnitude difference between the best-fit relations of the Sandage
\& Perelmuter (1991) and the Postman \& Lauer (1995) data; therefore,
the best-fit lines given in Table 2 of Paper I need to be made {\it
fainter} by the absolute value of the corrections listed in Table 3 of
Paper I. The resulting local lines carry an error in the $\langle SB
\rangle$ vertical position that varies between 0.04 mag and 0.50 mag
depending on log R (see Table 3 of Paper I).

{\it The effect for which we are searching is clearly seen in Figure
1.}  It is the depressed surface brightness at a given radius compared
with the surface brightness of early-type galaxies at zero
redshift. We interpret this depression to be the Tolman signal as
diluted by luminosity evolution.
 
The amount of the depression for each galaxy was calculated relative
to the zero-redshift line at a given radius. This is done by using the
equations for the zero-redshift lines from Tables 2 and 3 of Paper I
and subtracting the observed $\langle SB \rangle$ values for each
galaxy from these fiducial lines read at the same log R value. The
subsequent errors on this difference in mean surface brightness
include both the measurement error in the $\langle SB \rangle$ of the
galaxy and the uncertainties in the local line (see Tables 2--3 of
Paper I).  The result for Cl 1604+4321 is listed in Table 5 as the
mean depressions for the five Petrosian $\eta$ radii of 1.0, 1.3, 1.5,
1.7, and 2.0 using all of the galaxies in Table 2.

Column 2 of Table 5 shows the depression in magnitudes from the local
line which was calculated from a weighted average of the program
galaxies in Cl 1604+4321.  Column 3 of Table 5 gives the exponent,
$n$, in 2.5 log $(1 + z)^n$ for the observed mean $\Delta\langle SB
\rangle$ depression, which are listed as magnitudes in column 2. With
$z = 0.9243$, it follows that $n = 1.407 \times \Delta \langle SB
\rangle$ for Cl 1604+4321.

Column 4 of Table 5 shows the correction for luminosity evolution, in
magnitudes, that is required in order to make the Tolman signal of 2.5
log $(1 + 0.9243)^4 = 2.84$ mag, based on the argument in \S 1.4.3
that the true exponent must be 4 because of the black body
normalization of the relic radiation.  Therefore, column 4 is the
difference between $2.84$ mag and the listings in column 2.  Column 5
shows the exponent, $4 - n$, for the luminosity evolution, expressed
as $\Delta M_{evol} = 2.5~{\rm log}~(1+z)^{4-n}$ mag that is required
to make the true Tolman signal $(1 + z)^4$. Finally, column 6 shows
the number of galaxies in Cl 1604+4321 that are used in the averages.
     
Table 5 shows that the apparent Tolman signal in columns 2 and 3
dominates over the evolutionary requirements for luminosity evolution
in columns 4 and 5. The exponent $n$ varies from $3.43 \pm 0.23$ to
$2.48 \pm 0.25$ for $\eta$ radii from $\eta = 1.0$ to $2.0$ for Cl
1604+4321. 

\subsubsection{The Data in $I$ for the Two Clusters Cl 1324+3011 
($z = 0.7565$) and Cl 1604+4304 ($z = 0.8967$)}
 
Figure 2 shows the combined data in the $I$ band for the two clusters
observed with HST in the $F814W$ filter, reduced to the standard $I$
system and corrected for $K$ dimming by the values listed in Table
1. The $M$ and log R values listed in Tables 3 and 4 for these
clusters are calculated from the Mattig (1958) equations using the $m
- M$ and log $A$ values listed in Table 1 and the observed data in
Tables 5 and 6 of Paper III.  Filled circles indicate the galaxies in
Cl 1324+3011, while open circles indicate the galaxies in the
higher-redshift cluster Cl 1604+4304. Again, the different symbol
sizes separate the galaxies with measured redshifts from the assumed
cluster members based on photometric redshifts (see Paper III).

The differences between the observed $\langle SB \rangle$ magnitudes
and the zero-redshift calibration values at a given radius, given as a
weighted average over the galaxies in each cluster, are listed in
Tables 6--7 in the same format as for Table 5. Column 2 shows the mean
surface brightness depression from the zero-redshift fiducial line,
expressed in magnitudes.

Column 3 shows the exponent on $2.5~{\rm log}~(1 + z)$ to produce the
listed mean values of these surface brightness depressions.  Clearly,
$n = 1.635 \times \Delta \langle SB \rangle$ for Cl 1324+3011 ($z =
0.7565$) and $n = 1.439 \times \Delta \langle SB \rangle$ for Cl
1604+4304 ($z = 0.8967$). The exponent $n$ varies from $4.15 \pm 0.25$
to $3.25 \pm 0.25$ for Cl 1324+3011 and from $4.20 \pm 0.26$ to $3.29
\pm 0.30$ for Cl 1604+4304 for $\eta$ radii from 1.0 to 2.0.

\subsubsection{The Mean Tolman Signal From the Three Clusters} 

Tables 5--7 show the values of the mean depression in $\langle SB
\rangle$ and, therefore, the exponent $n$ are a strong function of
Petrosian radii $\eta$, being measured larger at smaller $\eta$
values. This trend is a direct result of the finite angular resolution
of the WFPC2 point-spread-function described in Paper II (Lubin \&
Sandage 2001a).  Specifically, for smaller values of $\eta$ we are not
accurately measuring the true $\eta(r)$ curve (see \S 2 of Paper
II). As a result, we {\it underestimate} the mean surface brightness
$\langle SB(r) \rangle$ and, therefore, {\it overestimate} the signal,
i.e.\ the depression from the zero-redshift relation. We have shown in
Paper II that, even for the smallest galaxies with half-light radii of
$0\farcs{25}$, the true $\eta(r)$ curve is reached only for $\eta
\gtrsim 1.8$. At these $\eta$ values, the error in the measured mean
surface brightness is less than 0.07 mag.

Therefore, in order to calculate our final numbers, we disregard the
data at $\eta$ values of less than 1.7 as undoubtedly unreliable at
the $> 0.1$ mag level. Consequently, we calculate the mean value of
the exponent $n$ averaged over the two $\eta$ values of 1.7 and 2.0
only.  We obtain a final value of the exponent $n$ in the $R$ band by
using the data from Cl 1604+4321. For the $I$ band, we note that the
resulting values of the exponent $n$ (as a function of $\eta$) are the
same within the errors for the two clusters observed in the $I$ band,
Cl 1324+3011 and Cl 1604+4304 (see Column 3 of Tables 5--7).
Therefore, we have combined the data for both clusters for the final
measurement in the $I$ band.

As discussed above, we calculate the values for the depression $\Delta
\langle SB \rangle$ (as represented by the exponent $n$) for the $R$
and $I$ bands from a weighted average of the data from $\eta = 1.7$
and 2.0. The results for $q_o = {1\over{2}}$ are

\begin{eqnarray}
\Delta \langle SB \rangle & = & 2.5~{\rm log}~(1 + z)^{2.59 \pm
0.17}~{\rm mag}~(R~{\rm Band}) \nonumber \\ 
& & \mbox{} 2.5~{\rm log}~(1 + z)^{3.37 \pm
0.13}~{\rm mag}~(I~{\rm Band}) 
\end{eqnarray}
	
\noindent for the apparent Tolman signal as compromised by luminosity
evolution. The required luminosity evolution would then be

\begin{eqnarray}
M_{evol} & = & 2.5~{\rm log}~(1 + z)^{1.41 \pm
0.17}~{\rm mag}~(R~{\rm Band}) \nonumber \\ 
& & \mbox{} 2.5~{\rm log}~(1 + z)^{0.63 \pm
0.13}~{\rm mag}~(I~{\rm Band}) 
\end{eqnarray}

\noindent We adopt equations (7) and (8) as our final values 
from the current experiment.
     
We must now test whether equation (8) is reasonable within the
independent discipline of luminosity evolution of the stellar content
using the precepts of population synthesis. In \S 4 we use the 1996
version of the Bruzual \& Charlot (1993) models to test for
compatibility between equation (8) and the stellar-population
synthesis models.  There is, of course, now a vast literature on
population synthesis models. We have only used the Bruzual \& Charlot
(1993) models here. It will eventually be important to test for
compatibility using different codes, such as earlier papers by
Guiderdoni \& Rocca-Volmerange (1987, 1988), Rocca-Volmerange \&
Guiderdoni (1988), and Worthey (1994). 

\subsection{The Doubly Diluted Tolman Signal in the $\langle SB \rangle$, $M$
Correlation Diagram}

In the last section we chose to analyze the data in the $\langle SB
\rangle$, log R plane rather than in the $\langle SB \rangle$, $M$
plane as was done in Sandage \& Perelmuter (1991). As explained in \S
3.1, the present method is preferred because it minimizes the effect
of luminosity evolution, which, from the result in the last section,
must be present at the level of $2.5~{\rm log}~(1+z)^{1.41 \pm 0.17}$
mag in the $R$ band and $2.5~{\rm log}~(1+z)^{0.63 \pm 0.13}$ in the
$I$ band if the radius does not evolve over the look-back time and if
the expansion is real.  Nevertheless, it is of interest to examine the
data in the $\langle SB \rangle$, $M$ correlation diagram that is
doubly degenerate to the effects of luminosity evolution.

Figure 3 shows the data for Cl 1604+4321 in the $R$ photometric band
with the same symbols as in Figure 1. The data for the individual
galaxies are from Table 2 and, therefore, refer to $H_o = 50$ and $q_o
= {1\over{2}}$ in calculating $M$ from the redshift. The ordinate of
$\langle SB \rangle$, as before, is independent of both $H_o$ and
$q_o$ because it is given directly as observed, except that it has
been corrected for $K(R)$ dimming via Table 1.  

We also plot in Figure 3 the envelope lines for the zero-redshift
calibration. These relations have been calculated from the equations
of $\langle SB \rangle$ versus log R given in Table 2 of Paper I, as
modified by the non-linearity corrections, given in Table 3 of Paper
I, at radii smaller than log R $= 4.3$.  We use equation (11) of Paper
I, which gives $\langle SB \rangle = M + 5~{\rm log~R(pc)} + 22.815$,
to calculate the absolute magnitude $M$ from the $\langle SB \rangle$,
log R local relations.  The resulting envelope lines, which are
plotted in Figure 3, are the same to within the errors as the linear
least-squares fits to the Postman \& Lauer (1995) local data given in
Table 4 of Paper I {\it over the overlapping linear range}.

Figure 4 shows the $I$ band data for Cl 1324+3011 and Cl 1604+4304 as
taken from Tables 3 and 4. The envelope lines plotted in this figure
are the same as in Figure 3; however, they have been converted to the
$I$ band by $\langle R-I \rangle = 0.62$ (see above).

Two features of Figures 3 and 4 are noted. (1) The scatter is larger
than in the $\langle SB \rangle$, log R plane (Figures 1--2). (2)
Although the mean $\langle SB \rangle$ in both diagrams is depressed
relative to the zero-redshift calibration lines, the amount is less in
the $\langle SB \rangle$, $M$ plane than in the $\langle SB \rangle$,
log R plane. The first effect on the larger scatter in Figures 3--4 is
due to the larger scatter in the $\langle SB \rangle$, $M$ plane even
for zero redshift (see Figure 3 of Paper I). The obvious reason is
that errors in $M$ and/or $\langle SB \rangle$ move points {\it
perpendicular} to the slope of the correlation line. By contrast, the
error vectors in the $\langle SB \rangle$, log R plane in Figures 1
and 2 are nearly parallel to the correlation lines.

The second effect is due to the double degeneracy to luminosity
evolution in the $\langle SB \rangle$, $M$ plane. As said before,
luminosity evolution affects both coordinates.  Relative to the
zero-redshift line, an increase in the luminosity at high redshift due
to luminosity evolution moves a point brighter in $M$ and also
brighter in $\langle SB \rangle$ relative to that line.  Therefore, in
an observed $\langle SB \rangle$, $M$ diagram that contains luminosity
evolution, the depression from the zero-redshift line is diluted
twice, once in $\langle SB \rangle$ and once in $M$.

Because Figures 1--2 are more powerful than Figures 3--4 for this
reason of double degeneracy, we have made the calculations of the
signal, as represented by $n$, and the luminosity evolution, as
represented by $4-n$, only from the $\langle SB \rangle$, log R data
in Figures 1--2. Of course, the final answers from the analysis in
either representation must be the same because the $\langle SB
\rangle$ versus log R and the $\langle SB \rangle$ versus $M$ diagrams
are simply different representations of the same data, connected by
equation (11) of Paper I.

\subsection{The Tautalogical Absolute Magnitude Signal in the 
log R, $M$ Plane} 

Use of the Mattig (1958) equations for linear radius and absolute
magnitude guarantees that the combination of $\langle SB \rangle$,
$M$, and the linear radius, R, are consistent with the variation of
$\langle SB \rangle$ with redshift as $2.5~{\rm log}~(1+ z)^4$ mag.
Because the log R and $M$ values in Tables 2--4 are calculated using
the Mattig (1958) equations, the deviations of the ``observed'' $M$,
calculated with the Mattig equation, must be precisely $2.5~{\rm
log}~(1 + z)^{4-n}$. Hence, the observed deviations from the
zero-redshift lines prove nothing except to show how much the absolute
magnitude, $M$, must change in the look-back time to conform with the
${\rm log}~(1+z)^n + {\rm log}~(1+z)^{4-n} = {\rm log}~(1+z)^4$
relation.

In Figures 5 and 6, we plot the relation between log R and $M$ for the
$R$ band and $I$ band, respectively. The envelope lines are calculated
from the zero-redshift relations using the linear equations from Table
2, the non-linear corrections for small R from Table 3, and equation
(11) given in Paper I.  These figures do nothing more than demonstrate
that $M$ {\it must} become brighter with increasing redshift (due to
evolution) in order to make the true depression of $\langle SB
\rangle$ in Figures 1 and 2, corrected for evolution, equal to
$(1+z)^4$ if the standard model is correct.  From equation (8), we see
that the requirements for a precise agreement with the Tolman
prediction is that the mean magnitude deviation at constant log R must
be $\langle \Delta M \rangle = 0.39 \pm 0.08$ mag in the $I$ band for
Cl 1324+3011 ($z = 0.7565$), $\langle \Delta M \rangle = 0.44 \pm
0.09$ mag in the $I$ band for Cl 1604+4304 ($z = 0.8967$), and
$\langle \Delta M \rangle = 1.00 \pm 0.12$ mag in the $R$ band Cl
1604+4321 ($z = 0.9243$) if the Tolman signal, freed from luminosity
evolution, would be precisely $(1+z)^4$.  Calculating the average
difference at constant log R between the high-redshift data and the
zero-redshift relations plotted in Figures 5 and 6, we measure
precisely, by definition, these values.  The predictions for
luminosity evolution from spectral synthesis models with varying star
formation histories are made in the next section to compare with these
predictions from the Tolman test.

\section{Theoretical Luminosity Evolution Estimated Using Two Different
Methods}

\subsection{Elementary Estimate of the Correction for Passive Luminosity 
Evolution Due To Main Sequence Burn Down}

An early order-of-magnitude estimate of the expected change of the
luminosities of early-type galaxies in the look-back time was based on
the change of the turnoff luminosity in the HR diagram for an old,
coeval population where no new stars are formed after the initial
starburst. The method is interesting because of its simplicity.  It
led early on to the result that the passive luminosity evolution
correction was close to $\Delta M_{bol} \sim \Delta M_V = 2.5~{\rm
log}~(1 + z)$.  This result is given by the simplistic calculation of
direct main sequence burn-down, including corrections for the main
sequence luminosity function (Sandage 1961b, 1988). In addition, the
elaborate population synthesis models by Tinsley (1968, 1972a,b, 1976,
1977, 1980 and references therein) gave nearly the same result.

We show in the next section that nearly the same result is also given
by fitting the observed spectral energy distribution and the size of
the 4000\AA\ break with the star formation models of Bruzual \&
Charlot (1993).  The look-back times vary, of course, with the assumed
value of $q_o$. The calculated ages for galaxies with the observed
redshifts of the three clusters must agree with the ages from the
Bruzual \& Charlot models, at least approximately, if our story is to
have coherence. Anticipating the next section, we set out here the
``cosmological ages'' based on the redshift, the look-back time, and
the assumed values of the Hubble constant $H_o$ and $q_o$.  The
general case of the calculation of the look-back time for all values
of $q_o$ is solved elsewhere (Sandage 1961b). The special cases of
$q_o = 0$ and $q_o = {1\over{2}}$ are, of course, trivial. The ages,
$T$, at a given redshift are given by

\begin{equation}
T = {{2 H_o^{-1}}\over{3(1+z)^{3\over{2}}}},
\end{equation}

\noindent for $q_o = {1\over{2}}$ and

\begin{equation}
T = {{H_o^{-1}}\over{(1+z)}}, 
\end{equation}

\noindent for $q_o = 0$.

\noindent The case for $q_o = 1$ is more complicated and can be found
from the tables in Sandage (1961b).

Equation (9) gives ages of 4.82 Gyr, 4.30 Gyr, and 4.21 Gyr at the
time light left the three clusters with redshifts of 0.7565, 0.8967,
and 0.9243, respectively, for $q_o = {1\over{2}}$ and $H_o = 58$ km
s$^{-1}$ Mpc$^{-1}$.  For these calculations, we must use the real
value of $H_o$ (Sandage \& Tammann 1997; Theureau et al.\ 1997; Saha
et al.\ 1999; Sandage 1999; Parodi et al.\ 2000) rather than an
arbitrary value as in previous sections because we need real ages in
this section.

Equation (10) gives the ages at the time light left of 9.60 Gyr, 8.89
Gyr, and 8.76 Gyr, respectively, for the same clusters using $q_o =
0$.  The extreme case of $q_o = +1$, where the age of the universe is
only $0.571 \times H_o^{-1} = 9.63$ Gyr for $H_o = 58$, gives ages
when light left the three clusters of 3.77 Gyr, 3.32 Gyr, and 3.26
Gyr, respectively, using Table 3 of Sandage (1961b).

It is also interesting to calculate the absolute magnitudes of the
main sequence termination ($TO$) for these ages. The modern oxygen
enhanced evolutionary models by Bergbush \& VandenBerg (1992) for
[Fe/H] = 0 give bolometric magnitudes of

\begin{equation}
        M(bol,TO) = 2.564~{\rm log}~T(yr) - 21.644,
\end{equation}

\noindent as interpolated from their tables (Sandage 1993). Hence, the
main sequence absolute bolometric magnitudes for the $q_o =
{1\over{2}}$ look-back times for the three clusters are 3.18, 3.06 and
3.03 mag, respectively. The brightest turn off magnitudes are, of
course, for the $q_o = +1$ models because the ages when light left are
the smallest. These bolometric turnoff magnitudes for the $q_o = +1$
case are 2.91, 2.79, and 2.73 mag, respectively.

\subsection{The Expected Evolution From the Bruzual \& Charlot 
(1993) models}

To calculate the expected amount of luminosity brightening with
look-back time, we have chosen to use the stellar population synthesis
code of Bruzual \& Charlot (1993) because, as part of the original
photometric and spectral analyses of these clusters, Postman, Lubin \&
Oke (1998, 2001; hereafter PLO98 and PLO01) use the 1996 stellar
evolution models of Bruzual \& Charlot (hereafter BC96) to study the
star-formation histories and ages of the cluster galaxies.  The
authors find that the Bruzual \& Charlot models are ideal because they
can be used to generate absolute energy distributions over a broad
wavelength range, with a spectral resolution which is comparable to
the observed Keck spectra (see Oke, Postman \& Lubin 1998; PLO98;
PLO01).

The free parameters in the BC96 models include the mass function, the
metal abundance, and the star-formation rate.  Following PLO98 and
PLO01, we have chosen models with a Salpeter mass function (Salpeter
1955) and a maximum stellar mass of $125~{\rm M_\odot}$.  Comparisons
with models constructed with a Scalo luminosity function (Scalo 1986)
show no significant difference at the level of accuracy that can be
achieved with low-resolution Keck spectra (PLO98).  Based on the
metal-line equivalent widths and balmer jump strengths, PLO01
determine that models with assumed metallicities between $Z = 0.004$
(0.2 solar) and $Z = 0.020$ (solar) provide reasonable fits to the
data. Because the metallicities of the cluster galaxies are not
strongly constrained, we assume a solar abundance for our
calculations.

The most significant constraint on the BC96 models is the choice of
the star formation history.  The simplest model assumes that there is
a large, initial burst of star formation after which the galaxy fades
in accordance with passive stellar evolution models.  These model are
called ``ssp'' models by Bruzual \& Charlot.  The next simplest models
are those where star formation begins at $t=0.0$ and decreases
exponentially with a fixed time constant.  PLO98 and PLO01 refer to
these models as tau ($\tau$) models and have considered time constants
ranging from 0.2 to 20 Gyr. Any of these models can be used to
generate the expected broad-band ($BVRI$) AB magnitudes once the
galaxy redshift is specified (see e.g.\ Figure 7 of Paper III).

In order to characterize the observed spectral energy distribution
(SED) of each galaxy, PLO01 have used the slope, referred to as $b$,
of a linear least-squares fit to the measured AB magnitudes as a
function of log $\nu$ (see \S 2.1 of PLO01). The slope $b$ provides a
more robust indicator of the overall broad-band SED than any
individual color measure, although $b$ is strongly correlated with the
usual broad-band colors [e.g.\ $(V-R) \approx {b\over{13}}$]. The
early-type galaxies used in the present experiment are the reddest
galaxies in the clusters and, therefore, have the largest values of
slope $b$ with $b \gtrsim 10$ (see Tables 2--4 of PLO01). The best-fit
BC96 models to these galaxies are either the ssp model (which is
essentially equivalent to a $\tau = 0.2$ Gyr model) or a tau model
with $\tau \le 1.0$ Gyr (see Figure 6 of PLO01).  Therefore, we have
chosen these models to measure the range in luminosity evolution that
we expect between $z = 0$ and the cluster redshifts of $z = \{0.7565,
0.8967, 0.9243\}$, respectively.

For these calculations, we use the ages for each cluster derived in \S
4.1 for different values of $q_o$ and assume that the epoch of star
formation occurred at $z_{form} \gtrsim 2.5$.  This high formation
epoch is consistent with other observations of cluster early-type
galaxies (e.g.\ Arag\'on-Salamanca et al.\ 1993; Stanford, Eisenhardt
\& Dickinson 1995, 1998; Oke, Gunn \& Hoessel 1996; Ellis et al.\
1997; van Dokkum et al.\ 1998; de Propis et al.\ 1999).

Using these models to measure the luminosity brightening with
look-back time, we find, as expected, that the smallest luminosity
evolution is measured with the ssp model because the burst of star
formation is the shortest of the models that we are considering;
conversely, the largest luminosity evolution is measured with the
$\tau = 1.0$ Gyr model because of its extended burst of star
formation.  For a $\tau = 0.2$ Gyr model with $q_o = {1\over{2}}$, we
find that the theoretical luminosity evolution would be $\Delta
M_{evol} (I) = 0.80$ mag for Cl 1324+3011 at $z = 0.7565$, $\Delta
M_{evol} (I) = 0.99$ mag for Cl 1604+4304 at $z = 0.8967$, and $\Delta
M_{evol} (R) = 1.12$ mag for Cl 1604+4321 at $z = 0.9243$. These
values are $\sim 0.4$ mag larger in the $I$ band that what we measure
from our analysis (see equation 8). The same difference is found in a
comparison between the observations and the models for the brightest
cluster galaxy (BCG) in each of our program clusters. On average, the
BC96 models predict a BCG magnitude which is $\sim 0.4$ mag brighter
than observed (see \S 6.1 of PLO01).

If we include the full range of models and cosmologies, the amount of
luminosity evolution expressed as the exponent, $p$, in 2.5 log
$(1+z)^p$ varies between $p = 0.85 - 2.36$ in the $R$ band and $p =
0.76 - 2.07$ in the $I$ band, depending on the particular model and
$q_o$. For a given star-formation model, the amount of evolution
increases as $q_o$ increases. In magnitudes, this luminosity evolution
corresponds to $\Delta M_{evol} (I) = 0.49 - 1.25$ mag for Cl
1324+3011, $\Delta M_{evol} (I) = 0.53 - 1.44$ mag for Cl 1604+4304,
and $\Delta M_{evol} (R) = 0.61 - 1.68$ mag for Cl 1604+4321.

The results of the theoretical calculations agree well with other
measures of luminosity evolution for early-type galaxies at $z \sim
0.5 - 1$. Depending on the particular passband and cosmological
parameters, the absolute luminosities of field and cluster early-type
galaxies are brighter by approximately $1.0 \pm 0.5$ mag at redshifts
of $ z \gtrsim 0.5$ (e.g.\ Im et al.\ 1996; Oke, Gunn \& Hoessel 1996;
Schade et al.\ 1996, 1999; Smail et al.\ 1997; van Dokkum et al.\
1998; Postman, Lubin \& Oke 1998, 2001).

Clearly, the amount of evolution that we require to make $n=4$ (see
equation 8) is fully consistent, within the errors, with the range of
theoretical expectations determined above. Therefore, we assert that
we have either (1) detected the evolutionary brightening directly from
the $\langle SB \rangle$ observations on the assumption that the
Tolman effect exists, or (2) confirmed that the Tolman test for the
reality of the expansion is positive provided that the theoretical
luminosity correction for evolution is real.

Our conclusions are fully consistent with the two previous attempts to
perform the Tolman Test using primarily ground-based data at more
moderate redshifts. Specifically, Sandage \& Perelmuter (1991)
completed an identical analysis comparing local ellipticals with the
first-ranked elliptical galaxies in clusters at redshifts up to $z
\sim 0.6$. Later, Pahre et al.\ (1996) used the Kormendy relation
between effective radius and mean surface brightness to compare
elliptical galaxies in the Coma cluster to those in Abell 2390 ($z =
0.23$) and Abell 851 ($z = 0.41$). Both studies found that the data
were fully consistent with the universal expansion assuming simple
models of passive evolution of elliptical galaxies.

\section{Sensitivity of the Conclusions to the Assumption that $q_o = 0$
and $q_o = +1$}

We need now to estimate the sensitivity of the results in \S 3 to the
value of $q_o$. The sensitivity arises from the differences in the
values of the linear radii calculated from the angular radii and the
redshift for different values of $q_o$.  Different linear radii
entered into the log R, $\langle SB \rangle$ diagnostic diagrams of
Figures 1 and 2 will give different displacements from the
zero-redshift local envelope lines and, therefore, a different Tolman
signal. We can estimate these differences by calculating the change in
the linear radii made by changing the value of $q_o$ and calculating
the effect on the Tolman signal using the zero-redshift envelope lines
for different $\eta$ values from Tables 2 and 3 of Paper I.

The Mattig (1958) equations for linear radius R $= f(q_o,z,H_o)$ and
absolute magnitude $M$ are summarized elsewhere (Sandage 1988; 1995,
eqs.\ 3.9 \& 4.5) for any $q_o$ and arbitrarily large $z$ values.
They will not be repeated here. Table 8 shows the result of using
these equations to calculate the necessary parameters for the $q_o =
0$ and $+1$ models.

The distance moduli listed in columns 3 and 7 are for $H_o = 50$. The
log $A$ values to convert log (angular radii) to log (linear radii)
are given in columns 4 and 8 via equations (5) and (6) of \S 2.1. The
differences in absolute magnitude and log R relative to the $q_o =
{1\over{2}}$ case are listed in columns 5--6 and 9--10 for the $q_o =
0$ and the $q_o = +1$ cases, respectively. The magnitude and log R
differences relative to the $q_o = {1\over{2}}$ case can be used to
convert the $M$, log R data in Tables 2--4 to the other $q_o$
cosmologies.

Calculation of the changes in the Tolman signal for different $q_o$
values is made by multiplying the change in log R from columns 6 and
10 of Table 8 by the slopes in Table 2 of Paper I.  These slopes vary
between 3.0 and 3.5 depending on the $\eta$ value. For our most
reliable values of $\eta = 1.7$ and 2.0, the average slope is
3.1. Multiplying the $\Delta$ Log R values by 3.1 shows that the
Tolman signal will be made smaller for the $q_o = 0$ case relative to
that for $q_o = {1\over{2}}$. That is, there will be a smaller
deviation from the local line.  The required evolutionary correction
is made larger by 0.18 mag for Cl 1324+3011 at $z = 0.7565$, 0.23 mag
for Cl 1604+4304 at $z = 0.8967$, and 0.22 mag for Cl 1604+4321 at $z
= 0.9243$. This difference translates into changing the exponent for
the Tolman signal and the evolutionary corrections in equations (7)
and (8) for the $q_o = {1\over{2}}$ case to $n = 2.28 \pm 0.17$ and
$4-n = 1.72 \pm 0.17$ for the $R$ band and $n = 3.06 \pm 0.13$ and
$4-n = 0.94 \pm 0.13$ for the $I$ band.

The opposite sense is required for the $q_o = +1$ case because Table 8
shows that the log R values are smaller than for the $q_o =
{1\over{2}}$ case by $\{0.057, 0.064, 0.065\}$ dex for $z = \{0.7565,
0.8967, 0.9243\}$, respectively. Using smaller log R values in Figures
1 and 2 gives a larger Tolman signal (in magnitudes) by 3.1 times the
differences in log R which are listed in Table 8. Consequently, the
measured luminosity evolution is smaller by 0.11 mag for Cl 1324+3011
at $z = 0.7565$, 0.12 mag for Cl 1604+4304 at $z = 0.8967$, and 0.15
mag for Cl 1604+4321 at $z = 0.9243$. These changes correspond to a
final Tolman signal of $n = 2.81 \pm 0.17$ and $n = 3.55 \pm 0.13$ in
the $R$ and $I$ band, respectively. The corresponding evolutionary
requirement is, of course, $4-n = 1.19 \pm 0.17$ and $0.45 \pm 0.13$
for the $R$ and $I$ band, respectively.

The conclusion is that the Tolman test as performed here is sensitive
to the value of $q_o$ at the less than 23\% level. However, as argued
in Paper I, it is not so severe as to make the test degenerate. Our
result is that a Tolman signal exists, as modified by evolution, at
the level of $(1 + z)^n$ with $n$ lying between $2.28 - 2.81~(\pm
0.17)$ in the $R$ band and between $3.06 - 3.55~(\pm 0.13)$ in the $I$
band, depending on $q_o$. Therefore, the correction for luminosity
evolution is $(1 + z)^p$ where $p$ lies between $1.72 - 1.19~(\pm
0.17)$ in the $R$ band and between $0.94 - 0.45~(\pm 0.13)$ in the $I$
band, again depending on $q_o$.  The required evolutionary correction
to make the real Tolman signal equal to $(1 + z)^4$ is well within the
errors of the requirements of Bruzual \& Charlot spectral synthesis
models, i.e.\ approximately 2.5 log $(1 + z)^p$ where $p \approx 0.8$
to 2.4 in the $R$ band and $p \approx 0.7$ to 2.1 in the $I$ band, as
$q_o$ changes from $0$ to $+1$.  Therefore, we conclude that the
Tolman prediction is verified and that the expansion is real.

We note that recent studies of high-redshift Type Ia supernovae, the
Cosmic Microwave Background, and the statistics of gravitational
lenses suggest that the universe is flat with a non-zero cosmological
constant, $\Lambda$ (see e.g.\ Kochanek 1996; Helbig et al.\ 1999; de
Bernadis et al.\ 2000; Pryke et al.\ 2001; Riess et al.\ 2001 and
references therein).  Currently, the preferred world model is
$\Omega_M \approx 0.35$ and $\Omega_{\Lambda} \approx 0.65$.  In this
cosmology, the log R values are almost identical to the empty-universe
($q_o = 0$) case; they are larger by only 0.01 dex or less at the
redshifts of our three clusters.  In addition, all flat-universe
models with $0 \le \Omega_{\Lambda} \le 0.65$ give log R values which
lie within the range that we have calculated for the $\Omega_{\Lambda}
= 0$ cosmologies (see Table 8).  Consequently, our conclusions about
the universal expansion are still robust for these $\Lambda$
cosmologies.

In the next section, we show that the predictions of the ``tired
light'' speculation is not verified at the definitive level of better
than 10 $\sigma$.

\section{The Tired Light Model Compared with the Observations}

In contrast with the standard model with the Mattig equations, there
is no metric theory of how distances and magnitudes are measured in a
``tired light'' model.  Therefore, we must guess at a reasonable
equation for distance. We adopt a discussion that is given elsewhere
(Sandage 1995, \S 4.3) and use an equation for ``coordinate'' distance
in a flat space which is 

\begin{equation}
D = \left ({c\over{H_o}} \right)~{\rm ln}(1 + z).
\end{equation}

Because the universe is not expanding in the model, the distance
``now'' in equation (12) is also the distance when light left. This,
of course, is the crucial point. In the expanding case, the distance
at the present epoch must be divided by $(1 + z)$ to give the distance
when light was emitted. It is this latter distance in the expanding
case that, when multiplied by the angular radius, fixes the linear
radius.

With $H_o = 50$, the linear radius that corresponds to an observed
angular radius (in arcsec) in the ``tired light'' case is
   
\begin{equation}
   {\rm log~R(pc)} = 4.464 + {\rm log}~[{\rm ln} (1 + z)] + {\rm log}~\theta",
\end{equation}

\noindent using equation (12). Hence, the $A$ term, as defined in
equation (5), becomes

\begin{equation}
          {\rm log}~A = 4.464 + {\rm log}~[{\rm ln} (1 + z)].
\end{equation}

The log $A$ values calculated in this way are listed in column 12 of
Table 8 for the three high-redshift clusters. These values, compared
with the log $A$ values for the $q_o = {1\over{2}}$ case, give the
increase in the log R values that must be entered in the $\langle SB
\rangle$, log R diagnostic diagram. For example, for Cl 1324+3011 ($z
= 0.7565$) the log R values must be made larger by 0.305 dex relative
to the $q_o = {1\over{2}}$ values listed in Table 3.
     
The magnitudes must also be changed because the distance moduli are
different than those in the fiducial $q_o = {1\over{2}}$ case. The
absolute magnitude calculation for the tired light model follows
from the expected theoretical relation that 

\begin{equation}
	l = {L\over{4\pi D^2(1+z)}},
\end{equation}

\noindent where only one power of $(1+z)$ for the ``energy effect'' is
required, rather than two powers of $(1+z)$ in the Roberston (1938)
equation in the standard theory (cf.\ Sandage 1995, eq.\ 2.1). Hence,

\begin{equation}
	m-M = 2.5~{\rm log}~(1+z) + 5~{\rm log}~[{\rm ln}(1+z)] +43.89,
\end{equation}

\noindent for $H_o = 50$.

It is can be shown from equation (16) that the magnitude--redshift
relation for ``tired light'' is the same to within a few hundredths of
a magnitude with the case for $q_o = +1$ using the Mattig
equation. This is true even for redshifts as large as $z = 1$. Column
13 of Table 8 confirms this statement, seen by the fact that the
entries in column 9 for the $q_o = +1$ case are very close to those in
column 13 for ``tired light.'' However, these magnitude changes are
academic here in using the $\langle SB \rangle$, log R diagnostic
diagram because no corrections for the magnitude differences need to
be applied to the $\langle SB \rangle$ values. These are {\it
observed} $\langle SB \rangle$ values (only corrected for $K$
dimming); hence, as emphasized before, they are independent of all
cosmologies.

With the changes to Tables 2--4 for log R that are listed in column 14
of Table 8, we can enter Figures 1 and 2 for the ``tired light'' case
in order to measure the depression from the local upper envelope
calibration. The result, not shown but which the reader can recover
using the $\Delta$ log R values given in Table 8 together with Tables
2--4, is that a measurable $\langle SB \rangle$ depression at these
larger log R values is again present. It is, of course, smaller than
for the expanding case because of the larger log R values. But, by how
much?  The crucial question is whether it is so much smaller as to
conform to a depression of only 2.5 log $(1 + z)$ mag when the
correction for luminosity evolution is also applied.

We have analyzed the data for the three clusters in the same way as in
\S 3; however, we now use the correct log R values required by the
``tired light'' conjecture, using the recipes in Table 8. Expressing
the result of the $\langle SB \rangle$ depression from the
zero-redshift fiducial line in the $\langle SB \rangle$, log R diagram
as the exponent, $n$, in 2.5 log $(1 + z)^n$ gives the weighted mean
of $n = 1.61 \pm 0.13$ for the $R$ band and $n = 2.27 \pm 0.12$ for
the $I$ band. As described in \S3, we have used the results from the
most reliable $\eta$ values of $\eta = 1.7$ and $2.0$. The resulting
exponents are {\it too large} by approximately 5 and 10 $\sigma$,
respectively, compared to the $n=1$ prediction of the ``tired light''
scenario. Consequently, to produce coherence with the ``tired light''
model, we require negative luminosity evolution in the look-back time,
i.e.\ {\it galaxies must be fainter in the past}. No feasible model of
stellar evolution can produce such luminosity evolution with time.

In fact, just as in the expanding case, positive luminosity evolution
must have occurred in the look-back time because there is no way in
the tired light model to prevent the stellar content of galaxies from
evolving during the look-back time.

A static model where the redshift is not due to expansion is not the
same as a steady state model where the mean parameters of galaxies,
averaged over an ensemble of galaxies, are required to be the same at
all distances and at all times. For a steady state to exist, despite
the evolution of the stellar content of individual galaxies, requires
that there must be young and old galaxies in every volume of space and
at every cosmic time such that the mean age is the same at all
distances and times. This requires continuous galaxy formation at the
same rate at all cosmic times in order to maintain a constant mean age
everywhere, always.

However, such steady state models are not the same as static models
where the redshift is due to an unknown physical cause, either at the
source or in the intervening light path from the source to the
observer as originally postulated by Zwicky (1929). In fact, the
steady state models proposed by Bondi, Gold, and Hoyle are truly
expanding models where the redshift is due to the expansion. They are
not static models.  Furthermore, a early proof that steady state
models cannot be correct was the demonstration of the failure of the
predicted steady state color distribution, with its required mixture
of ages, to match the observed color distribution of early-type
galaxies (Sandage 1973).

Hence, as in the present paper, evolutionary corrections to magnitudes
must be applied to both the expanding and the static models in making
the present test. The supposed degeneracy of the Tolman test due to
the identity claimed by Moles et al.\ (1998) of the surface brightness
effect in both the expanding and a static (tired light) case is not
correct. Their error is due to a category mistake by confusing static
and steady-state models.  What Moles et al.\ have done is to combine a
static model with a steady state model. This is a higher order
departure from the standard expanding model than we have considered
here and is not the test we have made.  In any case, as stated above,
a steady state model can again be disproved by the color argument
(Sandage 1973). Hence, even the higher order model proposed by Moles
et al.\ cannot be correct on this ground alone.
     
The result of the present paper is that a static model, where the
redshift is due to an unknown physical cause, fails the surface
brightness test by a large factor. Such a model requires luminosity
evolution in the look-back time, just as in the expanding case. Based
on the analysis in \S4, we find that a good approximation for the
amount of increase in luminosity at the epoch of light emission is 2.5
log $(1 + z)^p$ mag, where $p > 0.7$.  Applying this correction to the
``tired light'' analysis gives the intrinsic ``tired light''
prediction for the corrected exponents of $> 2.31~(\pm 0.13)$ and
$>2.87~(\pm 0.12)$ for the $R$ and $I$ bands, respectively.  Each
value is more than 10 $\sigma$ from the required exponent of 1.0 if
the ``tired light'' scenerio were correct. We take this to be a
definitive proof that the hypothesis of non-expansion does not fit the
surface brightness data.

\section{Systematic Uncertainties in the Experiment : How can the Present 
Result be Improved?}

There are two systematic uncertainties in the present experiment.
Although neither of them are severe enough to jeopardize the results
presented in \S3 and \S5 that the expansion is real, each can be
overcome by more data with expanded boundary conditions on the
parameters compared to the data used here.

First, the principal uncertainty at small radii (log R $< 4.0$) is the
position of the zero-redshift fiducial line in the $\langle SB
\rangle$, log R diagram, relative to which the $\langle SB \rangle$
depressions for high-redshift galaxies are compared. The Postman \&
Lauer (1995) data (Figure 2 of Paper I) do not extend to radii smaller
then log R(pc) $= 4.0$.  Their data are confined to the first ranked
cluster early-type galaxies. They do not sample into the luminosity
function of each cluster to provide data for smaller galaxies. We have
extended the data to smaller radii with the sample in Sandage \&
Perelmuter (1991) to generate a non-linear correction at small radii
to the best-fit linear equations to the Postman \& Lauer (1995) data
(see Table 3 of Paper I). However, the Sandage \& Perelmuter (1991)
data are also only for the first few, brightest cluster galaxies,
again not going far into the fainter part of the luminosity function.
Hence, although Table 3 of Paper I gives our adopted extension to
radii as small as log R(pc) $= 3.3$, the uncertainties are large and
can be reduced by a more complete study of the $\langle SB \rangle$,
log R relation for fainter and smaller galaxies at low redshift.

Second, the three clusters studied here are near the faint end of the
distribution of absolute magnitude of first ranked galaxies in, for
example, the sample of first ranked cluster galaxies whose data are
listed by Kristian, Sandage \& Westphal (1978). The mean of the
distribution of absolute magnitude in the $R$ band for the Kristian et
al.\ sample is $\langle M_R \rangle = -24.5$, whereas the first-ranked
early-type galaxies in the three clusters studied have absolute
magnitudes of $M_R = -23.7$ for Cl 1604+4321, $-23.7$ for Cl
1604+4304, and $-24.2$ for Cl 1324+3011 (see PLO01). (Note again that
the system of $R_J$ magnitudes used by Kristian et al.\ is 0.25 mag
brighter than the Cape $R$ system used here). The galaxies in the
three clusters studied here have fainter absolute magnitudes and
smaller radii than the average local clusters, exacerbating the
problem described above.  Richer high-redshift clusters with brighter
and, therefore, larger first-ranked galaxies are known. A study of the
HST data from such clusters will improve the present Tolman test. We
suspect that the present experiment is only the beginning of similar
work that will be done in the coming years with HST on such clusters
as they are discovered and observed.

\acknowledgments 

We are grateful to Mark Postman and J.B.\ Oke for their permission to
use part of their extensive Keck and HST database for this paper and
the previous papers in the series. LML would like to thank Chris
Fassnacht for his essential material aids to this paper. AS is
grateful for a conversation with James Peebles concerning the proof of
the Tolman $(1 + z)^4$ factor from the exact normalization of the
observed Planck black body background radiation in the COBE
experiment, as discussed in \S 1.4.3.

LML was supported through Hubble Fellowship grant HF-01095.01-97A from
the Space Telescope Science Institute, which is operated by the
Associated Universities for Research in Astronomy, Inc.\ under NASA
contract NAS 5-26555. AS acknowledges support for publication from
NASA grants GO-5427.01-93A and G-06459.01-95A for work that is related
to data taken with the {\it Hubble Space Telescope}.

\clearpage

\begin{deluxetable}{rccccc}
\tablewidth{0pt}
\tablenum{1}
\tablecaption{Adopted $K$-corrections, Distance Moduli, and $A$ Factor for 
the Linear Radii using $q_o = {1\over{2}}$, $H_o = 50$}
\tablehead{
\colhead{Cluster} &
\colhead{$z$} &
\colhead{$K(R) \pm \Delta K(R)$} &
\colhead{$K(I) \pm \Delta K(I)$} &
\colhead{$m-M$} &
\colhead {log $A$} \\
\colhead{} &
\colhead{} &
\colhead{(mag)} &
\colhead{(mag)} &
\colhead{} &
\colhead{}}
\startdata
1324+3011 & 0.7565 & \nodata & $0.71 \pm 0.03$ & 43.57 & 3.910 \\
1604+4304 & 0.8967 & \nodata & $0.80 \pm 0.07$ & 43.97 & 3.924 \\
1604+4321 & 0.9243 & $1.89 \pm 0.06$ & \nodata & 44.04 & 3.926 \\
\enddata
\end{deluxetable}

\clearpage
\voffset +1.00in

\begin{deluxetable}{cccccccccccccccc}
\rotate
\scriptsize
\tablewidth{0pt}
\tabletypesize{\scriptsize}
\tablenum{2}
\tablecaption{The $M$, $\langle SB \rangle$, Log R Data for Cl 1604+4321 [$R$ Band; $q_o = {1\over{2}}$; $H_o = 50$]}
\tablehead{
\colhead{Gal} &
\multicolumn{3}{c}{$\eta = 1.0$} &
\multicolumn{3}{c}{$\eta = 1.3$} &
\multicolumn{3}{c}{$\eta = 1.5$} &
\multicolumn{3}{c}{$\eta = 1.7$} &
\multicolumn{3}{c}{$\eta = 2.0$} \\
\colhead{\#} &
\colhead{$M$\tablenotemark{a}} &
\colhead{$\langle SB \rangle$\tablenotemark{a}} &
\colhead{Log R(pc)} &
\colhead{$M$\tablenotemark{a}} &
\colhead{$\langle SB \rangle$\tablenotemark{a}} &
\colhead{Log R(pc)} &
\colhead{$M$\tablenotemark{a}} &
\colhead{$\langle SB \rangle$\tablenotemark{a}} &
\colhead{Log R(pc)} &
\colhead{$M$\tablenotemark{a}} &
\colhead{$\langle SB \rangle$\tablenotemark{a}} &
\colhead{Log R(pc)} &
\colhead{$M$\tablenotemark{a}} &
\colhead{$\langle SB \rangle$\tablenotemark{a}} &
\colhead{Log R(pc)}}
\startdata
  23\tablenotemark{b}& $-23.13 \pm  0.10$ & 18.48 &  3.224& $-23.46 \pm  0.09$ & 19.16 &  3.409&  $-23.57 \pm  0.09$ & 19.49 &  3.493& $-23.65 \pm  0.08$ & 19.81 &  3.571 & $-23.78 \pm  0.08$ & 20.48 &  3.729 \\
  29\tablenotemark{b}& $-23.11 \pm  0.10$ & 20.35 &  3.568& $-23.28 \pm  0.09$ & 20.71 &  3.674&  $-23.37 \pm  0.09$ & 20.97 &  3.742& $-23.44 \pm  0.08$ & 21.23 &  3.805 & $-23.56 \pm  0.08$ & 21.84 &  3.955 \\
  36& $-22.46 \pm  0.10$ & 20.24 &  3.420& $-22.82 \pm  0.09$ & 20.97 &  3.632&  $-22.94 \pm  0.08$ & 21.34 &  3.727& $-23.05 \pm  0.08$ & 21.74 &  3.809 & $-23.15 \pm  0.08$ & 22.28 &  3.966 \\
  38& $-22.63 \pm  0.10$ & 20.21 &  3.451& $-22.87 \pm  0.09$ & 20.71 &  3.583&  $-22.97 \pm  0.08$ & 21.02 &  3.666& $-23.06 \pm  0.08$ & 21.39 &  3.767 & $-23.20 \pm  0.08$ & 22.02 &  3.918 \\
  44\tablenotemark{b}& $-22.47 \pm  0.09$ & 21.06 &  3.586& $-22.84 \pm  0.09$ & 21.76 &  3.790&  $-22.95 \pm  0.09$ & 22.10 &  3.881& $-23.04 \pm  0.08$ & 22.40 &  3.959 & $-23.12 \pm  0.08$ & 22.82 &  4.058 \\
  46& $-22.30 \pm  0.10$ & 21.20 &  3.581& $-22.63 \pm  0.09$ & 21.89 &  3.778&  $-22.77 \pm  0.08$ & 22.27 &  3.879& $-22.86 \pm  0.08$ & 22.62 &  3.966 & \nodata            &\nodata&\nodata \\
  49& $-21.95 \pm  0.10$ & 20.07 &  3.303& $-22.56 \pm  0.09$ & 21.33 &  3.647&  $-22.71 \pm  0.09$ & 21.77 &  3.770& $-22.80 \pm  0.08$ & 22.10 &  3.853 & $-22.88 \pm  0.08$ & 22.46 &  3.930 \\
  57& $-22.15 \pm  0.10$ & 21.65 &  3.636& $-22.58 \pm  0.09$ & 22.50 &  3.887&  $-22.79 \pm  0.09$ & 23.05 &  4.029& $-22.88 \pm  0.08$ & 23.39 &  4.132 & $-22.92 \pm  0.08$ & 23.58 &  4.171 \\
  58& $-21.88 \pm  0.10$ & 20.38 &  3.345& $-22.24 \pm  0.09$ & 21.11 &  3.547&  $-22.45 \pm  0.09$ & 21.72 &  3.707& $-22.67 \pm  0.08$ & 22.54 &  3.913 & $-22.82 \pm  0.08$ & 23.25 &  4.083 \\
  65& $-22.10 \pm  0.10$ & 19.83 &  3.278& $-22.40 \pm  0.09$ & 20.44 &  3.441&  $-22.50 \pm  0.08$ & 20.74 &  3.515& $-22.57 \pm  0.08$ & 21.01 &  3.590 & $-22.64 \pm  0.08$ & 21.43 &  3.667 \\
  78\tablenotemark{b}& $-21.53 \pm  0.10$ & 20.70 &  3.336& $-21.86 \pm  0.09$ & 21.40 &  3.529&  $-22.00 \pm  0.09$ & 21.80 &  3.643& $-22.14 \pm  0.08$ & 22.30 &  3.759 & $-22.17 \pm  0.08$ & 22.48 &  3.801 \\
 112\tablenotemark{b}& $-21.29 \pm  0.10$ & 20.27 &  3.219& $-21.60 \pm  0.09$ & 20.94 &  3.383&  $-21.70 \pm  0.08$ & 21.27 &  3.477& $-21.79 \pm  0.08$ & 21.62 &  3.568 & $-21.89 \pm  0.08$ & 22.12 &  3.667 \\
 115\tablenotemark{b}& $-21.16 \pm  0.09$ & 21.09 &  3.325& $-21.55 \pm  0.10$ & 21.87 &  3.556&  $-21.64 \pm  0.08$ & 22.14 &  3.618& $-21.71 \pm  0.08$ & 22.44 &  3.718 & $-21.84 \pm  0.08$ & 22.95 &  3.860 \\
 144\tablenotemark{b}& $-21.01 \pm  0.10$ & 20.74 &  3.246& $-21.37 \pm  0.09$ & 21.49 &  3.461&  $-21.48 \pm  0.09$ & 21.82 &  3.536& $-21.58 \pm  0.08$ & 22.21 &  3.632 & $-21.69 \pm  0.08$ & 22.78 &  3.767 \\
\enddata
\tablenotetext{a}{Values have been corrected with the appropriate $K(R) = 1.89$ correction (see Table 1).}
\tablenotetext{b}{Member galaxy selected based on its photometric redshift (see Brunner \& Lubin 2000).}
\end{deluxetable}

\clearpage
\voffset +1.00in

\begin{deluxetable}{cccccccccccccccc}
\rotate
\scriptsize
\tablewidth{0pt}
\tabletypesize{\scriptsize}
\tablenum{3}
\tablecaption{The $M$, $\langle SB \rangle$, Log R Data for Cl 1324+3011 [$I$ Band; $q_o = {1\over{2}}$; $H_o = 50$]}
\tablehead{
\colhead{Gal} &
\multicolumn{3}{c}{$\eta = 1.0$} &
\multicolumn{3}{c}{$\eta = 1.3$} &
\multicolumn{3}{c}{$\eta = 1.5$} &
\multicolumn{3}{c}{$\eta = 1.7$} &
\multicolumn{3}{c}{$\eta = 2.0$} \\
\colhead{\#} &
\colhead{$M$\tablenotemark{a}} &
\colhead{$\langle SB \rangle$\tablenotemark{a}} &
\colhead{Log R(pc)} &
\colhead{$M$\tablenotemark{a}} &
\colhead{$\langle SB \rangle$\tablenotemark{a}} &
\colhead{Log R(pc)} &
\colhead{$M$\tablenotemark{a}} &
\colhead{$\langle SB \rangle$\tablenotemark{a}} &
\colhead{Log R(pc)} &
\colhead{$M$\tablenotemark{a}} &
\colhead{$\langle SB \rangle$\tablenotemark{a}} &
\colhead{Log R(pc)} &
\colhead{$M$\tablenotemark{a}} &
\colhead{$\langle SB \rangle$\tablenotemark{a}} &
\colhead{Log R(pc)}}
\startdata
   9& $-23.23 \pm  0.07$ & 19.11 &  3.425& $-24.03 \pm 0.05$ & 20.75 &  3.905 & $-24.46 \pm  0.04$ & 21.87 &  4.294&   \nodata          &\nodata&\nodata & \nodata            &\nodata&\nodata \\
  11& $-23.21 \pm  0.07$ & 19.89 &  3.574& $-23.87 \pm 0.05$ & 21.21 &  3.964 & $-24.11 \pm  0.04$ & 21.82 &  4.136& $-24.18 \pm  0.04$ & 22.04 &  4.222 & \nodata            &\nodata&\nodata \\
  12& $-23.31 \pm  0.06$ & 20.27 &  3.667& $-23.54 \pm 0.05$ & 20.73 &  3.802 & $-23.63 \pm  0.05$ & 20.97 &  3.866& $-23.68 \pm  0.04$ & 21.15 &  3.915 & $-23.73 \pm  0.04$ & 21.39 &  3.968 \\
  18& $-22.69 \pm  0.07$ & 19.65 &  3.424& $-23.05 \pm 0.05$ & 20.39 &  3.638 & $-23.20 \pm  0.05$ & 20.81 &  3.749& $-23.32 \pm  0.05$ & 21.24 &  3.859 & $-23.46 \pm  0.04$ & 21.95 &  4.029 \\
  21& $-22.45 \pm  0.08$ & 19.36 &  3.327& $-22.91 \pm 0.06$ & 20.31 &  3.603 & $-23.08 \pm  0.05$ & 20.81 &  3.734& $-23.20 \pm  0.05$ & 21.26 &  3.843 & $-23.29 \pm  0.04$ & 21.69 &  3.952 \\
  26& $-22.24 \pm  0.07$ & 20.20 &  3.446& $-22.88 \pm 0.05$ & 21.44 &  3.816 & $-23.02 \pm  0.05$ & 21.82 &  3.912& $-23.12 \pm  0.05$ & 22.16 &  3.995 & $-23.17 \pm  0.04$ & 22.41 &  4.070 \\
  29& $-22.47 \pm  0.06$ & 20.36 &  3.519& $-22.83 \pm 0.05$ & 21.08 &  3.742 & $-23.00 \pm  0.04$ & 21.55 &  3.851& $-23.25 \pm  0.04$ & 22.45 &  4.087 & $-23.32 \pm  0.04$ & 22.81 &  4.174 \\
  30\tablenotemark{b}& $-22.42 \pm  0.07$ & 20.52 &  3.546& $-22.61 \pm 0.05$ & 20.93 &  3.657 & $-22.72 \pm  0.05$ & 21.22 &  3.740& $-22.82 \pm  0.04$ & 21.62 &  3.841 & $-22.89 \pm  0.04$ & 21.95 &  3.918 \\
  40& $-21.78 \pm  0.07$ & 19.42 &  3.229& $-22.19 \pm 0.05$ & 20.28 &  3.447 & $-22.37 \pm  0.05$ & 20.81 &  3.588& $-22.49 \pm  0.04$ & 21.28 &  3.702 & $-22.58 \pm  0.04$ & 21.72 &  3.811 \\
  55\tablenotemark{b}& $-21.43 \pm  0.07$ & 20.50 &  3.348& $-21.69 \pm 0.05$ & 21.04 &  3.502 & $-21.78 \pm  0.05$ & 21.31 &  3.571& $-21.83 \pm  0.04$ & 21.52 &  3.619 & $-21.91 \pm  0.03$ & 21.94 &  3.710 \\
  59& $-21.60 \pm  0.08$ & 19.73 &  3.245& $-21.89 \pm 0.05$ & 20.30 &  3.395 & $-21.99 \pm  0.04$ & 20.57 &  3.465& $-22.04 \pm  0.04$ & 20.81 &  3.526 & $-22.11 \pm  0.04$ & 21.16 &  3.605 \\
  69& $-21.48 \pm  0.04$ & 19.24 &  3.109& $-21.71 \pm 0.06$ & 19.76 &  3.256 & $-21.80 \pm  0.04$ & 20.05 &  3.327& $-21.86 \pm  0.04$ & 20.29 &  3.382 & $-21.92 \pm  0.03$ & 20.61 &  3.455 \\
  74\tablenotemark{b}& $-21.27 \pm  0.07$ & 21.28 &  3.470& $-21.57 \pm 0.05$ & 21.91 &  3.648 & $-21.72 \pm  0.05$ & 22.34 &  3.764& $-21.92 \pm  0.04$ & 23.03 &  3.938 & $-21.99 \pm  0.04$ & 23.35 &  4.016 \\
\enddata
\tablenotetext{a}{Values have been corrected with the appropriate $K(I) = 0.71$ correction (see Table 1).}
\tablenotetext{b}{Member galaxy selected based on its photometric redshift (see Brunner \& Lubin 2000).}
\end{deluxetable}

\clearpage
\voffset +1.00in

\begin{deluxetable}{cccccccccccccccc}
\rotate
\scriptsize
\tablewidth{0pt}
\tabletypesize{\scriptsize}
\tablenum{4}
\tablecaption{The $M$, $\langle SB \rangle$, Log R Data for Cl 1604+4304 [$I$ Band; $q_o = {1\over{2}}$; $H_o = 50$]}
\tablehead{
\colhead{Gal} &
\multicolumn{3}{c}{$\eta = 1.0$} &
\multicolumn{3}{c}{$\eta = 1.3$} &
\multicolumn{3}{c}{$\eta = 1.5$} &
\multicolumn{3}{c}{$\eta = 1.7$} &
\multicolumn{3}{c}{$\eta = 2.0$} \\
\colhead{\#} &
\colhead{$M$\tablenotemark{a}} &
\colhead{$\langle SB \rangle$\tablenotemark{a}} &
\colhead{Log R(pc)} &
\colhead{$M$\tablenotemark{a}} &
\colhead{$\langle SB \rangle$\tablenotemark{a}} &
\colhead{Log R(pc)} &
\colhead{$M$\tablenotemark{a}} &
\colhead{$\langle SB \rangle$\tablenotemark{a}} &
\colhead{Log R(pc)} &
\colhead{$M$\tablenotemark{a}} &
\colhead{$\langle SB \rangle$\tablenotemark{a}} &
\colhead{Log R(pc)} &
\colhead{$M$\tablenotemark{a}} &
\colhead{$\langle SB \rangle$\tablenotemark{a}} &
\colhead{Log R(pc)}}
\startdata
   9& $-23.24 \pm  0.10$ & 20.59 &  3.651 &$-23.72 \pm  0.10$ & 21.55 &  3.938 & $-23.89 \pm  0.09$ & 22.03 &  4.062& $-23.98 \pm  0.09$ & 22.36 &  4.149 & $-24.04 \pm  0.09$ & 22.59 &  4.204 \\
  10\tablenotemark{b}& $-23.12 \pm  0.10$ & 20.90 &  3.687 &$-23.59 \pm  0.10$ & 21.82 &  3.963 & $-23.71 \pm  0.09$ & 22.16 &  4.058& $-23.80 \pm  0.09$ & 22.46 &  4.131 & \nodata                 &\nodata&\nodata \\
  13& $-23.17 \pm  0.10$ & 20.00 &  3.525 &$-23.41 \pm  0.10$ & 20.50 &  3.669 & $-23.51 \pm  0.09$ & 20.80 &  3.746& $-23.59 \pm  0.09$ & 21.10 &  3.821 & $-23.68 \pm  0.09$ & 21.53 &  3.924 \\
  34\tablenotemark{b}& $-22.06 \pm  0.10$ & 20.98 &  3.495 &$-22.35 \pm  0.09$ & 21.60 &  3.674 & $-22.50 \pm  0.09$ & 22.03 &  3.785& $-22.58 \pm  0.09$ & 22.32 &  3.866 & $-22.64 \pm  0.09$ & 22.60 &  3.930 \\
  46\tablenotemark{b}& $-22.01 \pm  0.11$ & 20.40 &  3.378 &$-22.25 \pm  0.10$ & 20.90 &  3.513 & $-22.34 \pm  0.09$ & 21.17 &  3.589& $-22.42 \pm  0.09$ & 21.46 &  3.661 & $-22.50 \pm  0.09$ & 21.90 &  3.761 \\
  50& $-21.83 \pm  0.11$ & 21.11 &  3.479 &$-22.18 \pm  0.10$ & 21.83 &  3.689 & $-22.39 \pm  0.09$ & 22.43 &  3.855& $-22.47 \pm  0.09$ & 22.73 &  3.919 & $-22.60 \pm  0.09$ & 23.38 &  4.077 \\
  58\tablenotemark{b}& $-21.72 \pm  0.10$ & 20.65 &  3.371 &$-21.94 \pm  0.09$ & 21.11 &  3.499 & $-22.03 \pm  0.09$ & 21.37 &  3.569& $-22.10 \pm  0.09$ & 21.63 &  3.632 & $-22.18 \pm  0.09$ & 22.03 &  3.725 \\
\enddata
\tablenotetext{a}{Values have been corrected with the appropriate $K(I) = 0.80$ correction (see Table 1).}
\tablenotetext{b}{Member galaxy selected based on its photometric redshift (see Brunner \& Lubin 2000).}
\end{deluxetable}

\clearpage

\begin{deluxetable}{cccccc}
\tablewidth{0pt}
\tablenum{5}
\tablecaption{Summary of the Tolman Signal and the Inferred Luminosity Evolution for Cl 
1604+4321 in the $R$ band using $q_o = {1\over{2}}$}
\tablehead{
\colhead{$\eta$} &
\colhead{$\Delta \langle SB \rangle$} &
\colhead{$n$} &
\colhead{$\Delta M_{evol}$} &
\colhead{$4-n$} &
\colhead {\#} \\
\colhead{} &
\colhead{(mag)} &
\colhead{} &
\colhead{(mag)} &
\colhead{} &
\colhead{} \\
\colhead{(1)} &
\colhead{(2)} &
\colhead{(3)} &
\colhead{(4)} &
\colhead{(5)} &
\colhead{(6)}}
\startdata
1.0	& $2.44 \pm 0.16$ & $3.43 \pm 0.23$ & $0.40 \pm 0.16$  & $0.57 \pm 0.23$  & 14 \\	
1.3	& $2.24 \pm 0.16$ & $3.15 \pm 0.23$ & $0.60 \pm 0.16$  & $0.85 \pm 0.23$  & 14 \\	
1.5	& $2.21 \pm 0.16$ & $3.11 \pm 0.23$ & $0.63 \pm 0.16$  & $0.89 \pm 0.23$  & 14 \\	
1.7	& $1.96 \pm 0.17$ & $2.76 \pm 0.24$ & $0.88 \pm 0.17$  & $1.24 \pm 0.24$  & 14 \\	
2.0	& $1.76 \pm 0.18$ & $2.48 \pm 0.25$ & $1.08 \pm 0.18$  & $1.52 \pm 0.25$  & 13 \\	
\enddata
\end{deluxetable}

\begin{deluxetable}{cccrrc}
\tablewidth{0pt}
\tablenum{6}
\tablecaption{Summary of the Tolman Signal and the Inferred Luminosity Evolution for Cl 1324+3011 in the $I$ band using $q_o = {1\over{2}}$}
\tablehead{
\colhead{$\eta$} &
\colhead{$\Delta \langle SB \rangle$} &
\colhead{$n$} &
\colhead{$\Delta M_{evol}$} &
\colhead{$4-n$} &
\colhead {\#} \\
\colhead{} &
\colhead{(mag)} &
\colhead{} &
\colhead{(mag)} &
\colhead{} &
\colhead{} \\
\colhead{(1)} &
\colhead{(2)} &
\colhead{(3)} &
\colhead{(4)} &
\colhead{(5)} &
\colhead{(6)}}
\startdata
1.0	& $2.54 \pm 0.15$ & $4.15 \pm 0.25$ & $-0.09 \pm 0.15$ & $-0.15 \pm 0.25$ & 13 \\	
1.3	& $2.33 \pm 0.14$ & $3.81 \pm 0.23$ & $0.12 \pm 0.14$  & $0.19 \pm 0.23$  & 13 \\	
1.5	& $2.33 \pm 0.13$ & $3.81 \pm 0.21$ & $0.12 \pm 0.13$  & $0.19 \pm 0.21$  & 13 \\	
1.7	& $2.13 \pm 0.14$ & $3.48 \pm 0.23$ & $0.32 \pm 0.14$  & $0.52 \pm 0.23$  & 12 \\	
2.0	& $1.99 \pm 0.15$ & $3.25 \pm 0.25$ & $0.46 \pm 0.15$  & $0.75 \pm 0.25$  & 11 \\	
\enddata
\end{deluxetable}

\begin{deluxetable}{cccrrc}
\tablewidth{0pt}
\tablenum{7}
\tablecaption{Summary of the Tolman Signal and the Inferred Luminosity Evolution for Cl 
1604+4304 in the $I$ band using $q_o = {1\over{2}}$}
\tablehead{
\colhead{$\eta$} &
\colhead{$\Delta \langle SB \rangle$} &
\colhead{$n$} &
\colhead{$\Delta M_{evol}$} &
\colhead{$4-n$} &
\colhead {\#} \\
\colhead{} &
\colhead{(mag)} &
\colhead{} &
\colhead{(mag)} &
\colhead{} &
\colhead{} \\
\colhead{(1)} &
\colhead{(2)} &
\colhead{(3)} &
\colhead{(4)} &
\colhead{(5)} &
\colhead{(6)}}
\startdata
1.0	& $2.92 \pm 0.18$ & $4.20 \pm 0.26$ & $-0.14 \pm 0.18$ & $-0.20 \pm 0.26$ & 7 \\	
1.3	& $2.73 \pm 0.17$ & $3.93 \pm 0.24$ & $0.05 \pm 0.17$  & $0.07 \pm 0.24$  & 7 \\	
1.5	& $2.72 \pm 0.17$ & $3.91 \pm 0.24$ & $0.06 \pm 0.17$  & $0.08 \pm 0.24$  & 7 \\	
1.7	& $2.43 \pm 0.18$ & $3.50 \pm 0.26$ & $0.35 \pm 0.18$  & $0.50 \pm 0.26$  & 7 \\	
2.0	& $2.29 \pm 0.21$ & $3.29 \pm 0.30$ & $0.49 \pm 0.21$  & $0.71 \pm 0.30$  & 6 \\	
\enddata
\end{deluxetable}

\clearpage
\voffset +1.00in

\begin{deluxetable}{cccccccccccccc}
\rotate
\scriptsize
\tablewidth{0pt}
\tabletypesize{\scriptsize}
\tablenum{8}
\tablecaption{Recipes to Convert the $q_o = {1\over{2}}$, $H_o = 50$ Tables 
for Log R and $m-M$ to the Other Cosmologies}
\tablehead{
\colhead{} &
\colhead{} &
\multicolumn{4}{c}{$q_o = 0$} &
\multicolumn{4}{c}{$q_o = 1$} &
\multicolumn{4}{c}{``Tired Light''} \\
\colhead{Cluster} &
\colhead{$z$} &
\colhead{$m-M$} &
\colhead{Log $A$} &
\colhead{$\Delta M$} &
\colhead{$\Delta$ Log R} &
\colhead{$m-M$} &
\colhead{Log $A$} &
\colhead{$\Delta M$} &
\colhead{$\Delta$ Log R} &
\colhead{$m-M$} &
\colhead{Log $A$} &
\colhead{$\Delta M$} &
\colhead{$\Delta$ Log R}\\
\colhead{(1)} &
\colhead{(2)} &
\colhead{(3)} &
\colhead{(4)} &
\colhead{(5)} &
\colhead{(6)} &
\colhead{(7)} &
\colhead{(8)} &
\colhead{(9)} &
\colhead{(10)} &
\colhead{(11)} &
\colhead{(12)} &
\colhead{(13)} &
\colhead{(14)}}
\startdata
Cl 1324+3011	& 0.7565 & 43.98 & 3.992 & 0.41 brighter & 0.082 larger & 43.28 & 3.853 & 0.29 fainter & 0.057 smaller & 43.26 & 4.215 & 0.31 fainter & 0.305 larger \\
Cl 1604+4304	& 0.8967 & 44.46 & 4.021 & 0.49 brighter & 0.097 larger & 43.66 & 3.860 & 0.31 fainter & 0.064 smaller & 43.62 & 4.270 & 0.35 faitner & 0.346 larger \\
Cl 1604+4321	& 0.9243 & 44.53 & 4.025 & 0.50 brighter & 0.100 larger & 43.73 & 3.861 & 0.32 fainter & 0.065 smaller & 43.68 & 4.280 & 0.36 fainter & 0.354 larger \\
\enddata
\end{deluxetable}

\clearpage

\begin{figure}
\epsscale{0.7}
\plotone{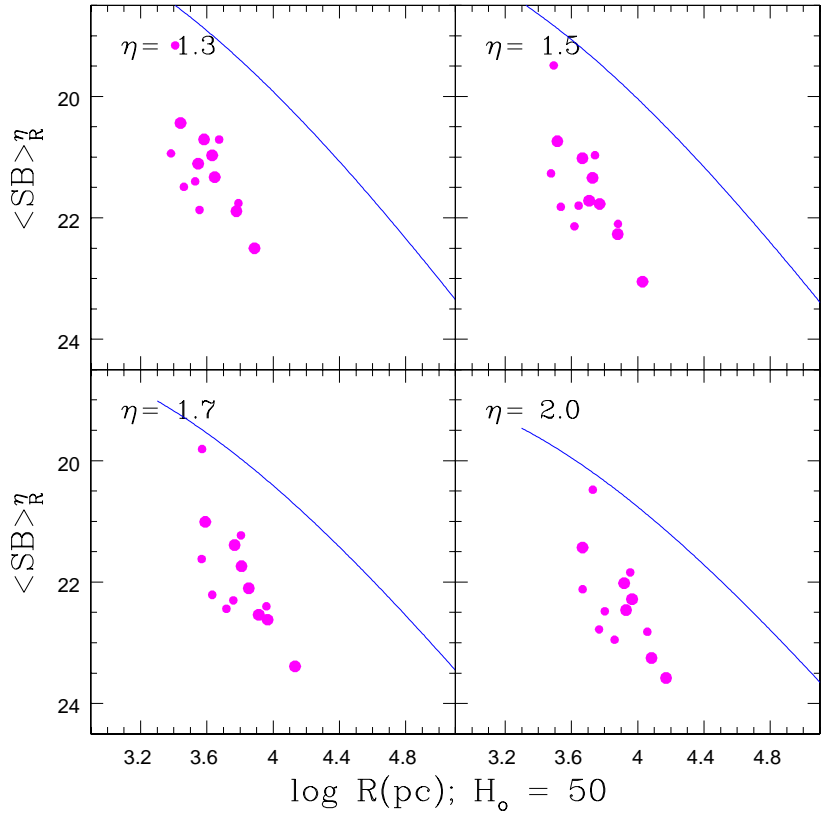}
\caption{Surface brightness measured in the $R$ band versus log linear
radius for the galaxies in Cl 1604+4321 ($z = 0.9243$) from the data
in Table 2. Large dots are for the galaxies in Table 2 with directly
measured redshifts. Small dots are cluster members inferred for their
photometric redshifts (see Paper III). The mean correlation for
zero-redshift early-type galaxies is shown by the lines taken from
Tables 2 and 3 of Paper I.}
\end{figure}

\begin{figure}
\epsscale{0.7}
\plotone{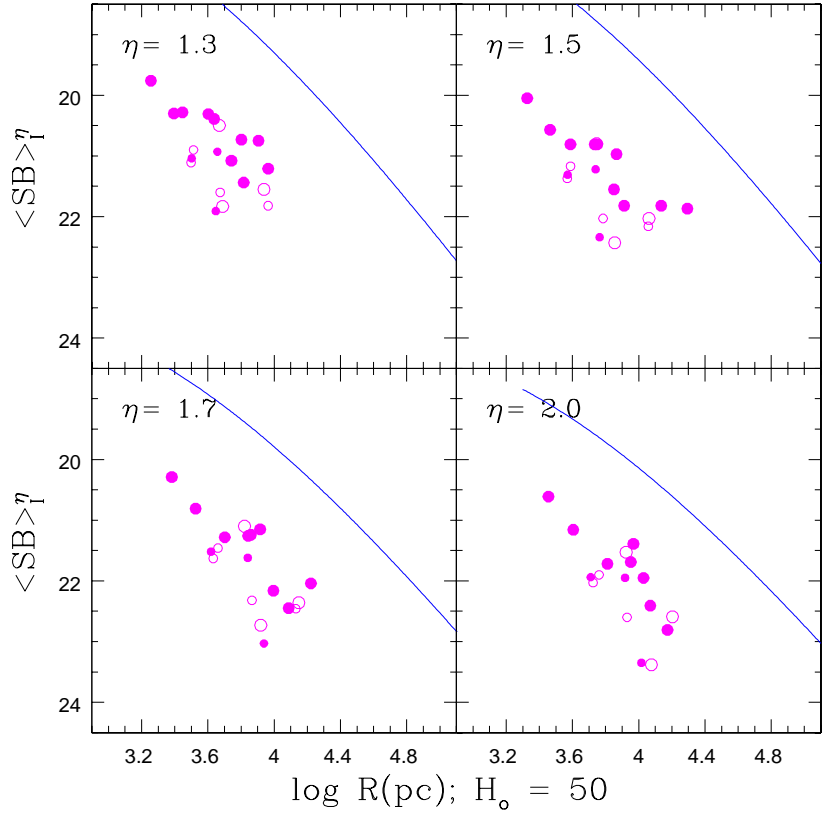}
\caption{Same as Figure 1 but for the $I$ band data for the two
clusters Cl 1324+3011 at $z = 0.7565$ (closed circles) and Cl
1604+4304 at $z = 0.8967$ (open circles) from the data in Tables 3 and
4. The lines for zero redshift are the same as in Figure 1 but shifted
to the $I$ band by $\langle R-I \rangle = 0.62$.}
\end{figure}

\begin{figure}
\epsscale{0.7}
\plotone{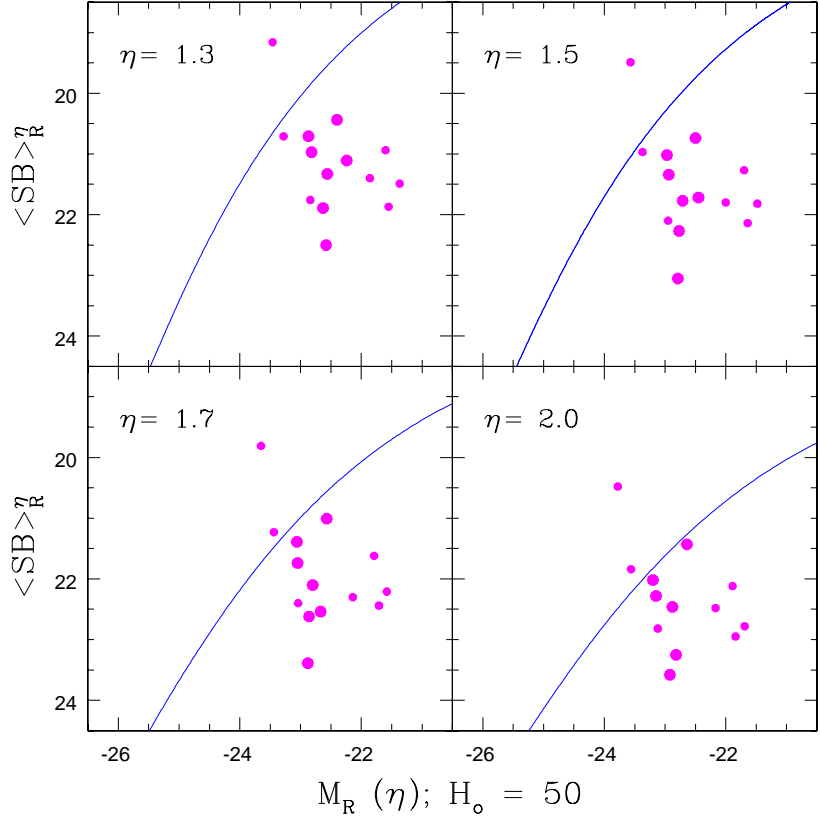}
\caption{The $M$, $\langle SB \rangle$ diagram for Cl 1604+4321 from
the data in Table 2 for the $H_o = 50$, $q_o = {1\over{2}}$ model. The
solid line is the zero-redshift calibration derived from the
zero-redshift log R, $\langle SB \rangle$ relation in Figure 1 and
equation (11) of Paper I. The symbols are the same as in Figure 1.}
\end{figure}

\begin{figure}
\epsscale{0.7}
\plotone{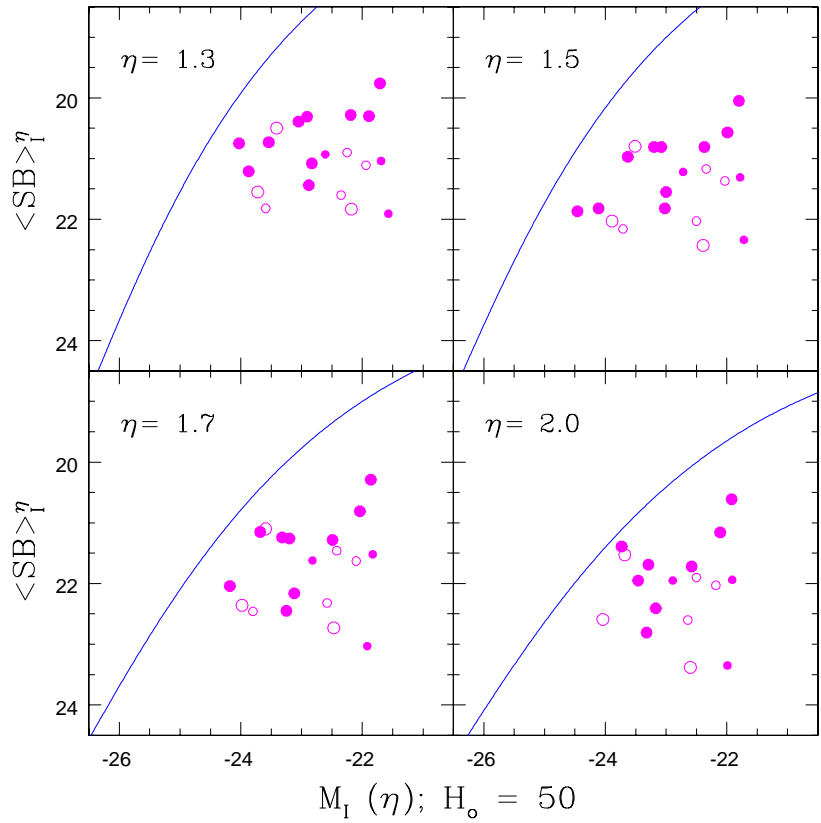}
\caption{Same as Figure 3 for the $M$, $\langle SB \rangle$ diagram
for the combined data for the clusters Cl 1324+3011 and Cl 1604+4304
from Tables 3 and 4. Closed and open circles are for Cl 1324+3011 and
Cl 1604+4304, respectively.}
\end{figure}

\begin{figure}
\epsscale{0.7}
\plotone{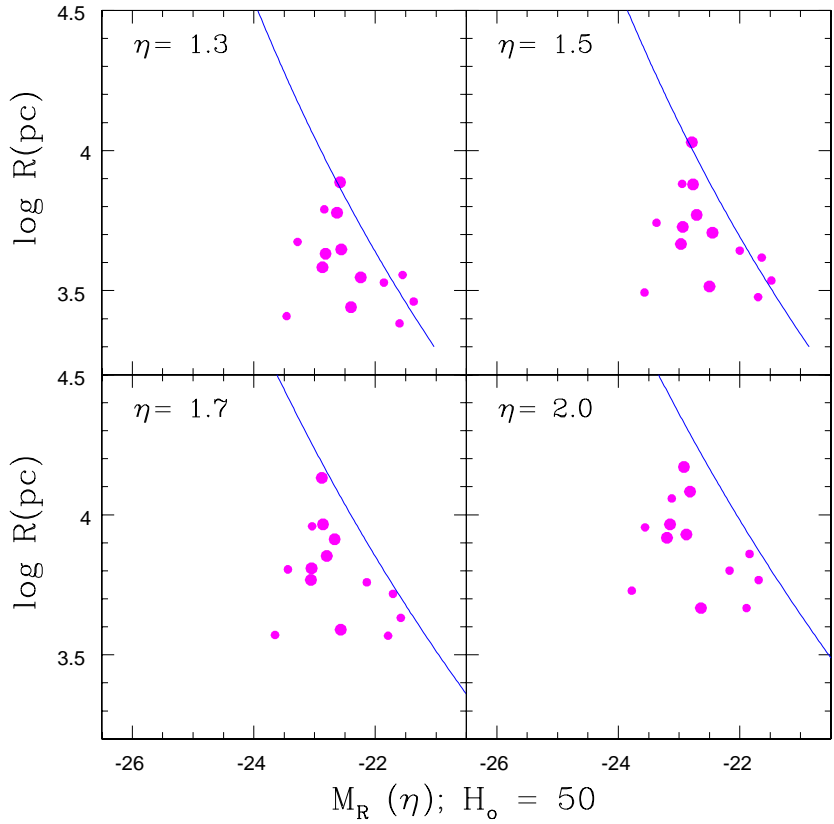}
\caption{The deviation of the absolute magnitude at a given linear
radius in the $R$ photometric band from the zero-redshift correlation
line for Cl 1604+4321. The solid line is the zero-redshift calibration
derived from the zero-redshift log R, $\langle SB \rangle$ relation in
Figure 1 and equation (11) of Paper I. The increase in absolute
luminosity that is required to preserve the $(1+z)^4$ Tolman factor is
the deviation toward brighter magnitudes seen relative to the
zero-redshift line. This is clearly due to luminosity evolution. To
preserve the Tolman factor requires that the evolutionary factor must
be $\langle \Delta M \rangle = 2.5~{\rm log}~(1.9243)^{1.41} = 1.00$
mag for the weighted mean of the $\eta$ values of 1.7 and 2.0. Because
of the tautology of the argument, this diagram must show this factor,
by necessity. The expectation from the theory of passive evolution is
set out in \S 4.}
\end{figure}

\begin{figure}
\epsscale{0.7}
\plotone{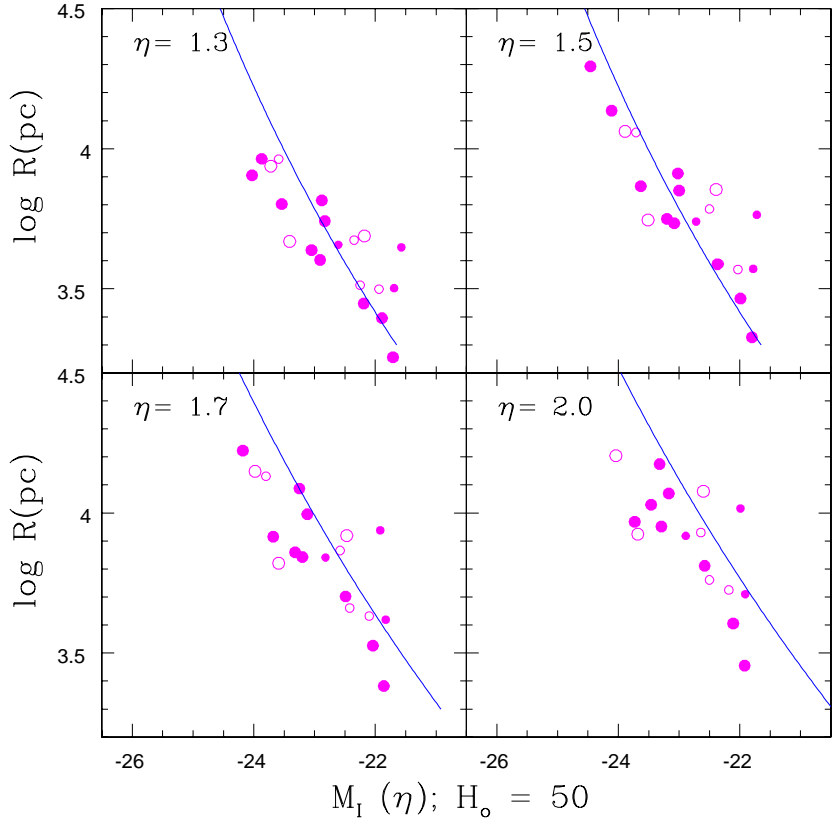}
\caption{Same as Figure 5 but in the $I$ photometric band for the two
clusters Cl 1324+3011 ($z = 0.7565$) and Cl 1604+4304 ($z = 0.8967$).
The tautological deviations in magnitudes from the zero redshift lines
are required to be $2.5~{\rm log}~(1.7565)^{0.63} = 0.39$ mag and
$2.5~{\rm log}~(1.8967)^{0.63} = 0.44$ mag, respectively, as the
weighted mean of the $\eta$ values of 1.7 and 2.0 if the expansion is
real.  These required evolutionary corrections are within the errors
of the independent calculation of the evolution correction from the
spectral synthesis calculations in \S 4 based on the observed colors.}
\end{figure}

\end{document}